# SOLVATION EFFECTS ON KINETICS OF METHYLENE CHLORIDE REACTIONS IN SUB- AND SUPERCRITICAL WATER: THEORY, EXPERIMENT, AND *AB INITIO* CALCULATIONS


P.A. Marrone[1], T.A. Arias[2], W.A. Peters[1], and J.W. Tester[1, *]

[1]Department of Chemical Engineering and Energy Laboratory
[2]Department of Physics
Massachusetts Institute of Technology
77 Massachusetts Ave.
Cambridge, MA  02139

617-253-3401  (tel.)
617-253-8013  (fax)
testerel@mit.edu  (e-mail)





*  To whom correspondence should be sent


## Abstract


The nature of the $CH_2Cl_2$ neutral/acidic hydrolysis reaction from ambient to supercritical conditions (25 to 600°C at 246 bar) is explored. Of primary interest is the effect of the changing dielectric behavior of the water solvent over this temperature range on this reaction. Experiments reveal that significant $CH_2Cl_2$ hydrolysis occurs under subcritical temperatures, while relatively little hydrolysis occurs under supercritical conditions. These trends cannot be explained by simple Arrhenius behavior. A combination of Kirkwood theory and *ab initio* modeling provides a means of successfully accounting for this behavior both qualitatively and quantitatively. The results show that increases in the activation energy and a changing reaction profile with a decreasing dielectric constant provide a mechanism for a slowing of the reaction at higher temperatures by as much as three orders of magnitude. These solvent effects are captured quantitatively in a correction factor to the Arrhenius form of the rate constant, which is incorporated into a global rate expression proposed for $CH_2Cl_2$ hydrolysis that provides good predictions of the experimental data.




# I. INTRODUCTION

In recent years, there has been much interest in the use, disposal, and environmental fate of chlorinated organic compounds. The past popularity of many of these halocarbons can be traced to their extensive use in industrial processes, most notably as solvents, extractants, cleaning agents, pesticides, and chemical intermediates. In the last decade, public concern has grown over their increasing presence and persistence in the environment, and their potential toxicity. For example, polychlorinated alkanes now account for a significant fraction of organic chemicals on the US EPA designated priority pollutant list.[1] Supercritical water oxidation (SCWO) is one innovative technology that has been shown to effectively destroy many types of organic compounds including these more refractory halocarbons.[2] At conditions above its critical point (374°C, 221 bar), water is relatively nonpolar due to diminished hydrogen bonding. Consequently, most nonpolar organic compounds and gases such as oxygen are completely miscible in supercritical water and rapidly react to form $CO_2$ and $H_2O$. Typically, greater than 99.99% destruction of organic species can be attained under conditions of 550-650°C, 250 bar, and residence times of 1 minute or less without formation of toxic by-products. In the oxidation of chlorinated compounds, Cl heteroatoms are ultimately converted to HCl. In the case of higher molecular weight and/or labile organic compounds, appreciable destruction can even be achieved under subcritical temperatures (wet oxidation) and without oxygen present (hydrolysis or pyrolysis conditions).

Over the last decade, *ab initio* calculations have been very successful in describing microscopic phenomena involving the rearrangements of chemical bonds such as when reactions occur in vacuum. Recently, it has been recognized that it is possible to couple such calculations with more coarse-grained macroscopic theories in unified multiscale approaches in order to study vastly more complex systems. Yet, exactly how this is to be done for different reactions remains an open question. A promising area is the coupling between a reacting molecular complex and a dielectric solvent.



Methylene chloride ($CH_2Cl_2$) in aqueous solution is a useful model system for this study because it has been a widely used solvent in industry and is a common environmental contaminant.[3]  On the molecular level, its dipole moment also has the potential to couple strongly with the dielectric water solvent.  Methylene chloride thus poses an interesting scenario for linking *ab initio* calculations to a macroscopic system.  Furthermore, study of $CH_2Cl_2$ also allows the details of C-Cl chemistry in sub- and supercritical water to be investigated using a comparatively simple substrate.  In an earlier paper,[4] data on $CH_2Cl_2$ conversion and the product spectrum and distribution were presented, along with a preliminary proposed reaction network.  There we noted a somewhat counterintuitive result: significant destruction of $CH_2Cl_2$ under subcritical temperatures (*i.e.*, in a section of the feed preheater with $T < 374°C$) and little additional reaction under generally more highly reactive supercritical conditions.  Our overall goal here is to explain this behavior.

Toward this end, the objectives of this paper are four-fold.  The first is to review the experimental data and trends observed, and to demonstrate that a simple Arrhenius form for the rate constant does not alone capture these trends.  The second is to establish that it is the influence of the changing dielectric properties of the solvent which dramatically slows the reaction at higher temperatures.  We will find that because of the rapidly changing dielectric constant, the transition state *itself* is highly sensitive to the solvent environment in this case.  The third objective is to develop a quantitative theory of the kinetics of such a reaction.  This will be accomplished by the use of a unique combination of Kirkwood theory, transition state theory, and *ab initio* quantum mechanical calculations.  Our final objective is to then use these results, along with experimental temperature profiles, to generate engineering-oriented global kinetic rate expressions for the hydrolysis reaction of $CH_2Cl_2$.  A more complete reaction network has also been developed for $CH_2Cl_2$ hydrolysis and its products based on our experimental data and studies in the literature and is presented by Marrone *et al.*[5]



Although there have been several studies of subcritical liquid and vapor hydrolysis of various chlorinated hydrocarbons,[6, 7] only a few have been carried out at supercritical conditions. Recent studies in supercritical water include that of Foy *et al.*[8] on hydrolysis and oxidation of trichloroacetic acid, trichloroethylene, and 1,1,1-trichloroethane; and of Houser and Liu [9] on hydrolysis of 1-chloro-3-phenylpropane, 2-chlorotoluene, and 4-chlorophenol in metal and glass ampoule batch reactors to explore catalytic effects. Some recent and relevant (although not supercritical) studies focusing specifically on reactions of $CH_2Cl_2$ include oxidative pyrolysis,[10] and several involving heterogeneous reactions such as catalytic hydrolysis,[11] catalytic oxidation in both dry and humid air,[12] and surface reduction with iron in the presence of water.[13]

## II.  SUMMARY OF EXPERIMENTAL DATA AND OBSERVATIONS

The heated portion of the experimental system (see Figure 1 and Marrone *et al.*[4] for more details) consisted of two separate preheater coils and a main tubular reactor heated in a fluidized sandbath. One preheater coil contained a pressurized aqueous feed solution of $CH_2Cl_2$ and another contained a pressurized $O_2$/water solution (for oxidation runs) or just pressurized pure deionized water (for hydrolysis runs). Each preheater coil was approximately 3 m in length, with only the lower 2.75 m heated (*i.e.*, submerged in the fluidized sand). The preheater tubing had an I.D. of 0.108 cm and a wall thickness of 0.025 cm, and was constructed of Hastelloy C-276. Both feeds entered their respective preheaters at ambient temperature and were heated to a specified operating temperature before reaching a mixing tee and the main reactor. Between 70 - 80% of the residence time in the preheater tubing was spent at subcritical temperatures. Measured mixing tee temperatures were usually within about 5°C of the sandbath temperature. Hydrolysis conditions always existed in the $CH_2Cl_2$ feed preheater coil in all experiments. For oxidation runs, oxidizing conditions existed only in the main reactor, after the point at which the aqueous $O_2$ and $CH_2Cl_2$ feed solutions were mixed.



The main reactor tube was 4.71 m in length, had a 0.171 cm I.D. and 0.232 cm wall thickness, and was constructed of Inconel 625. The main reactor was always kept at turbulent (and nearly plug) flow and isothermal conditions at a supercritical temperature between 450 and 600°C. The fluid temperature was measured just before exiting the main reactor at the top of the sandbath. The effluent fluid was then immediately quenched down to ambient temperature rapidly (*i.e.*, in < 1 s) in a shell and tube heat exchanger, followed by reduction of pressure to atmospheric.

Sandbath temperature, residence times, and feed concentrations were varied considerably over all of the experiments performed. Table 1 contains the complete range of experimental operating conditions for the preheater tubing and the main reactor. All experiments were isobaric at approximately 246 bar. Feed concentrations at the beginning of the preheater are values at ambient temperature, while at the entrance to the isothermal main reactor they correspond to the reactor operating temperature after adjusting for dilution due to the mixing of the two feed solutions. In all calculations, the concentrations of the $CH_2Cl_2$ and $O_2$ feed solutions were considered sufficiently dilute to permit the use of density values for pure water at the given temperature and pressure. The values for the initial $CH_2Cl_2$ feed concentrations and $[O_2]_o/[CH_2Cl_2]_o$ feed ratios in the main reactor differ from those cited in our earlier paper[4] because these current values now account for the considerable $CH_2Cl_2$ hydrolysis that occurred in the preheater tubing, as will be discussed shortly. Preheater tubing residence times, calculated from temperature profile models described below, and main reactor residence times are also provided in Table 1.

The major results and trends regarding $CH_2Cl_2$ reaction are summarized as follows:

- $CH_2Cl_2$ conversions ranged from 26±9% to 91±1% for pure hydrolysis experiments and 30±9% to 99.9±0.1% for oxidation experiments.
- $CH_2Cl_2$ conversions from hydrolysis runs were significant, and similar to those observed from oxidation runs under the same conditions at sandbath temperatures ≤ 525°C (see Figure 2).



- $O_2$ had a detectable effect on $CH_2Cl_2$ conversion only at sandbath temperatures > 525°C. At a constant sandbath temperature, varying the $O_2$ concentration over the range cited in Table 1 had little effect on $CH_2Cl_2$ conversion.

- Corrosion was observed in four separate $CH_2Cl_2$ feed preheater coils in a region up to at least 40 cm downstream of the point at which the tubing entered the sandbath. The most severe corrosion resulted in through-wall failure of each tube in a region between 7 and 29 cm. Tube life before failure ranged from 45 to 104 hrs of use at operating temperatures (see Mitton *et al.*[14, 15] for details). Because the only source of Cl⁻ (one of the most aggressive agents of corrosion) was $CH_2Cl_2$, HCl must have formed from $CH_2Cl_2$ breakdown sufficiently early in the preheater tubing to cause the corrosion to occur where it was observed near the beginning of the heated section. This evidence implies that reaction definitely occurred in this region of the preheater tube. This evidence alone, however, indicates nothing about the reaction rate further down the tube.

- Of the 23 hydrolysis experiments, two were performed at sandbath temperatures of 450°C and 575°C with the main reactor removed and replaced by a short 40 cm stainless steel tube which connected the mixing tee to the heat exchanger. Similar $CH_2Cl_2$ conversions were observed with and without the supercritical main reactor present (Table 2), even at comparable residence times. This evidence indicates that in fact only a very limited amount of further hydrolysis conversion occurred under supercritical conditions in the main reactor, and that the reaction rate was actually slower on average in the reactor than in the preheater tubing

The products that were detected and identified from both hydrolysis and oxidation reactions of $CH_2Cl_2$ included formaldehyde, hydrochloric acid, carbon monoxide, carbon dioxide, hydrogen gas, methanol, and trace amounts of methane, chloromethane, chloroform, trichloroethylene, and isomers of dichloroethylene.[4] The mean carbon mass balance over all runs was 99.9%, with a standard deviation of 6.6%. The mean chlorine mass balance over all runs was 100.0%, with a standard deviation of 5.7%. Despite the variety of products observed over the different



experimental operating conditions explored, only $CO_2$ and HCl were found under oxidizing conditions at a sandbath temperature of 600°C and a total residence time of 23 s (6 s in the presence of $O_2$). The prominent role of $O_2$ was in affecting the product spectrum and distribution, rather than the extent of $CH_2Cl_2$ conversion itself. For a further discussion of product-associated trends as they relate to the complete reaction network, see Marrone *et al.*[5]

## III. GENERAL NATURE OF THE CH₂Cl₂ HYDROLYSIS REACTION

The results summarized above indicate that substantial hydrolysis of $CH_2Cl_2$ occurred under subcritical, non-oxidative conditions in the preheater tubing. The appreciable and closely similar conversions between hydrolysis and oxidation experiments (Figure 2), the early corrosive failures of the $CH_2Cl_2$ feed preheater tubing, and the similarity in conversions observed with and without the supercritical main reactor in place (Table 2) all together suggest that for this apparatus most of the conversion of the initial $CH_2Cl_2$ feed occurred via hydrolysis in the subcritical zone of the preheater tubing. These data imply a rapid subcritical hydrolysis reaction that decreases in rate significantly as temperature increases into the supercritical regime. In order to explain this behavior, one must first consider the nature of the subcritical hydrolysis reaction.

In an earlier investigation, Fells and Moelwyn-Hughes[16] studied the hydrolysis kinetics of $CH_2Cl_2$ in liquid water from 80 to 150°C in a series of batch experiments conducted in Pyrex ampoules, under both neutral and basic conditions. Despite the lower temperatures and different experimental system, their kinetics results support the proposition that much of the $CH_2Cl_2$ breakdown observed in our apparatus occurred at subcritical temperatures.

The hydrolysis of $CH_2Cl_2$ can generally be classified as a nucleophilic substitution reaction. As is usually the case with methyl and methylene halides, the reaction mechanism is most likely of the single-step, bimolecular type ($S_N2$) rather than the two-step, unimolecular type ($S_N1$).[17, 18] The



products observed and intermediates postulated by Fells and Moelwyn-Hughes are consistent with this type of reaction. Under neutral conditions, water adds an OH group to the central carbon replacing one Cl, which combines with the extra $H^+$ to form HCl:

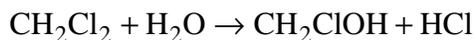

$$CH_2Cl_2 + H_2O \rightarrow CH_2ClOH + HCl$$

The resulting unstable species $CH_2ClOH$ then undergoes a much faster internal rearrangement, expelling another $H^+$ and $Cl^-$ to form HCHO:

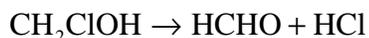

$$CH_2ClOH \rightarrow HCHO + HCl$$

The overall rate of reaction is dictated by the slower first substitution step.[16] Nucleophilic substitution reactions usually involve one or more charged or polar species (reactants or intermediates) and proceed via a polar mechanism. This probably accounts, in part, for why hydrolysis of $CH_2Cl_2$ occurred readily under subcritical temperatures, since the polar water solvent under these conditions could clearly support this type of reaction.

Although $H_2O$ and $OH^-$ are both possible nucleophiles for a hydrolysis reaction, $H_2O$ was more important under our operating conditions. While the $OH^-$ ion is a stronger nucleophile than $H_2O$ by about four orders of magnitude,[19] the concentration of $OH^-$ (formed only from dissociation of water in our experiments) was much less than that of the solvent water. At the most favorable conditions for $[OH^-]$, we have estimated the rate of neutral hydrolysis to still be greater than that of basic hydrolysis by at least a factor of $10^3$ (typically much higher) under the full range of experimental conditions employed. The presence of $H_2O$ as the dominant nucleophile is further supported by Schwarzenbach *et al.*,[20] who state that reactions of aliphatic halides in aqueous solution with $OH^-$ are unimportant at pH values less than about 10, which was well above the acidic pH of the present experiments.



# IV. FAILURE OF ARRHENIUS-FORM RATE EXPRESSION FOR HIGH TEMPERATURE $CH_2Cl_2$ HYDROLYSIS

Fells and Moelwyn-Hughes[16] proposed the following empirical, Arrhenius type correlation for the first order rate constant for neutral/acidic hydrolysis of $CH_2Cl_2$:

$$log\, k_{FMH} = 98.4408 - 29.66\, log\, T - 10597.3/T \qquad (1)$$

where $T$ is in Kelvins and $k_{FMH}$ in $s^{-1}$. Note that $k_{FMH}$ is almost of simple Arrhenius form, but includes an explicit temperature dependence in the pre-exponential factor via the $log\, T$ term. We explored the suitability of Eqn. 1 to predict the $CH_2Cl_2$ hydrolysis conversion in our experimental system. Note that this required extrapolation far outside the 80 to 150°C range for which the correlation was developed.

Following the approach of Holgate *et al.*,[21] a differential transient analysis of coupled heat transfer and kinetics was used to estimate the temperature and $CH_2Cl_2$ concentration variations in our experimental system. In the preheater tubing, the set of relevant differential equations is:

$$\frac{dT}{dz} = \frac{2\pi r_i U_i (T_{fsb} - T)}{\dot{m} C_p(T)} \qquad (2)$$

$$\frac{d[CH_2Cl_2]}{dz} = -\frac{R_{hyd}\, \pi r_i^2 \rho(T)}{\dot{m}} + \frac{[CH_2Cl_2]}{\rho(T)} \frac{d\rho(T)}{dT} \frac{dT}{dz} \qquad (3)$$

where the hydrolysis reaction rate $R_{hyd}$ (using Eqn. 1 for $k_{FMH}$) is given as:

$$R_{hyd} = \frac{d[CH_2Cl_2]}{dt} = -k_{FMH}(T)[CH_2Cl_2] \qquad (4)$$

Eqn. 2 is a differential heat balance where $z$ is the axial position in the tube, $r_i$ is the internal radius of the tube, $T_{fsb}$ is the sandbath temperature, $\dot{m}$ is the mass flow rate of the feed solution, $C_p$ is the heat capacity of the feed solution, and $U_i$ is the overall heat transfer coefficient based on inner surface area of the tube (see Appendix for details on the calculation of $U_i$ and experimental



temperature profiles). Because of the dilute feed solutions used, heat of reaction effects were negligible. Residence times $t$ in the preheater were calculated from Eqn. 2 by the following equation:

$$\frac{dt}{dz} = \frac{\rho(T)\pi r_i^2}{\dot{m}} \qquad (5)$$

where $\rho(T)$ is the density. In Eqn. 3, the first term accounts for chemical reaction, while the second term accounts for the effect of decreasing density that occurs as the reacting mixture is heated. In the isothermal main reactor, the concentration was determined directly from Eqn. 4, which is what Eqn. 3 reduces to under constant temperature. Eqns. 2, 3, 4, and 5 form a non-linear, coupled set that were solved simultaneously via a Runge-Kutta numerical integration technique applied over the entire length of the preheater tubing and main reactor to ultimately yield a value of $[CH_2Cl_2]$ at the end of the preheater and main reactor for each experiment. Because of the dilute nature of the feed, values for water density and other fluid properties in these calculations were taken from the NBS/NRC steam tables.[22]

Calculated $CH_2Cl_2$ concentration profiles using predicted temperature profiles and the Fells and Moelwyn-Hughes rate constant (Eqn. 1) are shown for our hydrolysis runs in Figure 3. Three sets of results are presented in Figure 3: the experimentally measured $CH_2Cl_2$ conversions at the end of the reactor system (also shown in Figure 2), values predicted from the model at the end of the preheater tubing, and values predicted from the model at the end of the isothermal main reactor. The data are plotted versus sandbath temperature, and calculated values of residence time in both the preheater and complete system (preheater and main reactor) are also shown to facilitate interpretation. Note that residence time does not scale linearly with sandbath temperature, as experimental mass flow rates were decreased with increasing sandbath temperature in our experiments to compensate for the increasing volumetric flow rates at higher temperatures. In all cases, the main reactor (supercritical) residence time was 6 s, which was always less than the calculated preheater residence time.



Figure 3 shows that the simple Arrhenius-type model of Fells and Moelwyn-Hughes has major shortcomings. Although the calculated values of conversion are of the same order of magnitude as the experimental values, the shape of the curves formed by the set of calculated conversion values (both at the end of the preheater tubing and at the end of the main reactor) clearly do not match the experimental trends. Note also that, in each case, almost half of the total conversion is predicted to occur in the supercritical main reactor, which contradicts the experimental observations (Table 2). These results are not necessarily unexpected, since the magnitude of the Fells and Moelwyn-Hughes rate constant (and indeed of any rate constant of Arrhenius form) should increase monotonically with increasing temperature. Thus, considering only temperature effects, one would in fact predict that the amount of total conversion would be very high in the supercritical main reactor, as opposed to the lower temperature preheater tubing, for a rate constant of this form. This model, however, is unable to capture the behavior shown by the experimental data and observations over our extended sub- and supercritical temperature range.

These discrepancies between experimental results and the Fells and Moelwyn-Hughes model suggest that the simple Arrhenius-type rate expression is failing to account for some important effect. Specifically, an additional temperature-dependent phenomenon must be offsetting the strong positive temperature effect of the Arrhenius term at high temperatures. Apart from the temperature, the one factor which varied significantly during the course of each experiment was the nature of the water solvent itself. Below we investigate the impact of solvation effects on the $CH_2Cl_2$ hydrolysis reaction rate.

## V. THE INFLUENCE OF THE SOLVENT ON REACTION RATES

When water is heated from ambient to supercritical conditions, it transforms from being a dense, strongly polar, hydrogen bonded liquid to a lower density, nonpolar fluid with gas-like diffusivity and viscosity.[2, 23, 24] Two quantities which are likely to affect reaction rates in



supercritical water are thus the density and dielectric nature of the solvent. The polar nature of the subcritical hydrolysis of $CH_2Cl_2$ makes it reasonable to suspect that the changing dielectric nature plays a key role. Below, we will present *ab initio*-based evidence that, in this case, the dielectric effect is two to three orders of magnitude more important than the density effect.

The literature provides strong precedent for noticeable solvent effects on reaction rates, particularly in the vicinity of the critical point of water. For example, the computer simulations of Gao and Xia[25] show that the rate of two reactions, the Claisen rearrangement of allyl vinyl ether and the $S_N2$ reaction of $NH_3$ and $CH_3Cl$ increase when conducted in aqueous solution as opposed to the gas phase. Researchers at the University of Texas at Austin have explored the solvation effects of supercritical water on a number of reactions via simulation and spectroscopy, including proton transfer reactions[26, 27] and the $S_N2$ reaction between $Cl^-$ and $CH_3Cl$.[28] Townsend *et al*.[29] observed parallel pyrolysis and hydrolysis mechanisms for the reaction of a number of coal model compounds in sub- and supercritical water, and found the product selectivity to be dependent upon the solvent. These trends were also seen in a similar study of nitroaniline explosive simulants.[30] In addition, Townsend *et al*.[29] found that the value of the rate constant for a few compounds was highly correlated to the dielectric constant of the water solvent. In a follow-up study, Huppert *et al*.[31] showed that the hydrolysis rate constant of guaiacol could be manipulated by changes in the solvent water density near the critical point (at a temperature of 383°C) by pressure variation and by adding salts. Similarly, in a study of 1-propanol dehydration in supercritical water, Narayan and Antal[32] observed a decrease in the acid-catalyzed rate constant by a factor of three when the water dielectric constant was increased by a proportional amount by an increase in pressure. Iyer and Klein[33] incorporated a dielectric constant dependent term in their correlation for the rate constant for butyronitrile hydrolysis to account for changes in the electrostatic nature of the solvent resulting from pressure variation. Finally, Habenicht *et al*.[34] observed improved fits in their reaction model when accounting for the influence of the solvent dielectric constant on the synthesis of ethyl tert-butyl ether from tert-butyl alcohol in liquid ethanol.



The qualitative picture of the dielectric solvent effect is that when the solvents tend to stabilize the transition state complex of a reaction (relative to the reactants), this enhances the reaction rate. For those reactions where the electric fields associated with the transition state are stronger, due to for instance a higher net charge or stronger dipole moment, a solvent with strong dielectric response or high polarity will tend to lower the energy of the transition state relative to the reactants, resulting in a faster reaction.[35] Conversely, for those reactions where the reactants have stronger electric fields than the transition state, a solvent with high dielectric constant will lower the total energy of the reactants more than that of the transition state, resulting in a higher activation energy and a slower reaction. The trends are opposite when the dielectric constant of the solvent is low. In the present case, the dielectric response of $H_2O$ decreases dramatically with increasing temperature near the critical point. We thus in principle expect the rate of an aqueous reaction with a highly polar transition state to slow down as the water is heated from subcritical to supercritical conditions, and for the reaction to accelerate if the reactants are more polar than the transition state.

For qualitative determination of solvent effects on rates of nucleophilic substitution reactions, Lowry and Richardson[35] have divided the family of nucleophilic substitution reactions into four main categories based on the electrical nature of the nucleophile and the leaving group. The neutral hydrolysis reaction of $CH_2Cl_2$, with the neutral nucleophile $H_2O$ and the charged leaving group $Cl^-$, falls under Type 2, which is the only category in which the transition state produces stronger electric fields than the isolated reactants. The basic hydrolysis of $CH_2Cl_2$, on the other hand, with charged nucleophile $OH^-$ and leaving group $Cl^-$, falls under Type 1, in which the electric fields of the isolated reactants are stronger than for the transition state. We thus expect a solvent medium of high dielectric constant to favor the neutral hydrolysis reaction, and a solvent medium of low dielectric constant to favor the basic hydrolysis reaction, relative to each other. As water goes through the critical point with increasing temperature, therefore, the rate of the neutral hydrolysis reaction is expected to be retarded due to the changing nature of the solvent. Although



one in principle would expect the rate of basic hydrolysis to accelerate under the same conditions, the greatly reduced water dissociation constant and low ion solubilities typical in supercritical water would likely result in a low $OH^-$ concentration available and prevent basic hydrolysis from becoming significant. The experimentally observed fast hydrolysis of $CH_2Cl_2$ under subcritical conditions but much slower reaction under supercritical conditions is thus consistent with the neutral hydrolysis mechanism.

## VI. THEORETICAL DEVELOPMENT

**Kirkwood Theory**  To account *quantitatively* for the solvent-induced slowing of the $CH_2Cl_2$ hydrolysis reaction in going from sub- to supercritical conditions, one starts by invoking transition state theory. This application is not straight-forward, however, as it requires accounting of the varying energy of solvation of the reactants and the transition state as the polar nature of the water reaction environment changes. Further complications arise as the conformation of the transition state itself depends on its molecular environment, an effect which can only be accounted for properly by an accurate quantum mechanical description of the reaction. The theoretical description which we use involves a unique combination of transition state theory, electrostatics and *ab initio* quantum calculations.

For a first attempt at such a combined theory, we shall take the picture originally developed by Kirkwood[36] as a reasonable starting point for computing the energy of solvation of a molecule, deferring refinements for future work. Kirkwood modeled the solvation energy of a species as the change in Gibbs free energy as the species is transferred from vacuum to a dielectric continuum with dielectric constant $\varepsilon$ equal to that observed experimentally for the solvent. To a first approximation, the charge distribution of the solute (represented by point charges $q_k$) is held fixed during this process and the dielectric constant of the solvent has the value $\varepsilon$ everywhere except inside a spherical volume of radius $r$ (representing the solute) from which the solvent is excluded by short range interactions with the solute (Figure 4). In this work, we follow the more common



practice of taking the dielectric constant within this sphere to be that of free space (*i.e.* $\varepsilon_i = 1$).[36, 37] With this approach, the energy of solvation is simply the energy required to create the charge distribution of the solute within the dielectric cavity. Solution to this standard electrostatic boundary value problem results in the following expression:

$$\Delta G_{solv,i} = \frac{Z^2 e^2}{8\pi\varepsilon_o r}\frac{1-\varepsilon}{\varepsilon} + \frac{\mu^2}{4\pi\varepsilon_o r^3}\frac{1-\varepsilon}{2\varepsilon+1} + \Kappa \tag{6}$$

where $\Delta G_{solv,i}$ is the energy of solvation for a species $i$, $Ze$ is the total charge of the solute, $\mu$ is its dipole moment, and $\varepsilon_o$ is the vacuum permittivity. Generally for species carrying a net charge, the first term dominates and the higher order terms are neglected.[38] For neutral but polar species, as are $CH_2Cl_2$ and $H_2O$ and their transition state, the first term is zero and the second dipolar term dominates the result. Below, we shall always truncate Eqn. 6 to just these two terms, again referring refinements to future work.

The subtle effects missed in this simple description of a solute molecule can in principle be treated correctly in a series of ever more demanding calculations (described in Section IX). Eqn. 6, however, is a reasonable approximation of at least semiquantitative reliability which we apply in this initial attempt to understand and characterize the reaction.

**Application to Transition State Theory** Transition state theory provides the connection between the solvation energy and reaction rate. The rate constant for two neutral molecular species A and B is

$$k = \frac{k_b T}{h} e^{-\Delta G^{\ddagger}/RT}, \tag{7}$$

where $k_b$ is Boltzmann's constant, $h$ is Planck's constant, $R$ is the universal gas constant, and $\Delta G^{\ddagger}$ is the standard Gibbs free energy of activation for the reaction of A and B to form the transition state complex $AB_{\ddagger}$. In this case, the $\Delta G^{\ddagger}$ can be divided into two separate terms for each species $i$ (reactant or product),



$$\Delta G^{\ddagger} = \sum_i \nu_i \left( G_i^o + \Delta G_{solv,i} \right). \tag{8}$$

Here $\nu_i$ is the stoichiometric coefficient for species $i$ according to the reaction, $G_i^o$ is the standard Gibbs free energy for a molecule of species $i$ in vacuum, and $\Delta G_{solv,i}$ is the solvation energy from the vacuum state for species $i$, as given in Eqn. 6. Note that the sum $\sum_i \nu_i G_i^o$ is just the standard state free energy change $\Delta G^o_{rxn}$ for the reaction in vacuum, which quantum mechanical calculations normally provide. Combining Eqns. 6, 7 and 8 gives:

$$\ln k = \ln k_o - \frac{N_A(1-\varepsilon)}{4\pi\varepsilon_o RT(2\varepsilon+1)} \left[ \frac{\mu_{\ddagger}^2}{r_{\ddagger}^3} - \frac{\mu_A^2}{r_A^3} - \frac{\mu_B^2}{r_B^3} \right] \tag{9}$$

where $N_A$ is Avogadro's number, $\mu_i$ and $r_i$ are the dipole moment and radius of species $i$, respectively, and $k_o$ is the rate constant for the reaction in vacuum. The second term on the right in Eqn. 9 containing the Kirkwood expression is the factor that "corrects" the magnitude of the rate constant for solvation effects. Note that, as expected from the previous qualitative arguments, the value of this Kirkwood correction factor increases the reaction rate as the solvent dielectric constant $\varepsilon$ increases, provided that the $\mu_{\ddagger}^2/r_{\ddagger}^3$ term for the transition state is greater than the sum of the corresponding terms for the reactants.

A more practical reference state than vacuum is the dielectric environment of ambient water ($\varepsilon = \varepsilon_a \approx 80$). To use this as the reference, we first compare the prediction of Eqn. 9 for such ambient conditions with the observable behavior near ambient temperature characterized by an Arrhenius form with pre-exponential factor $A$ and activation energy $E_a$:

$$\ln k = \ln A - \frac{E_a}{RT} \tag{10}$$

Equating Eqn. 9 evaluated at ambient conditions with Eqn.10 in order to solve for $k_o$, and then inserting the result back into Eqn. 9 yields:



$$\ln k = \ln A - \frac{1}{RT}\left(E_a + \frac{N_A}{4\pi\varepsilon_o}\left(\frac{(\varepsilon_a-1)}{(2\varepsilon_a+1)} - \frac{(\varepsilon-1)}{(2\varepsilon+1)}\right)\left[\frac{\mu_{\ddagger}^2}{r_{\ddagger}^3} - \frac{\mu_A^2}{r_A^3} - \frac{\mu_B^2}{r_B^3}\right]\right) \qquad (11)$$

This expression gives the reaction rate constant in terms of the activation energy and pre-exponential factor under ambient conditions, $E_a$ and $A$ respectively, and a Kirkwood-derived term which acts to slow the reaction as $\varepsilon$ decreases, provided that the transition state is more polar than the reactants.

In point of fact, the behavior of the reaction rate for the hydrolysis of $CH_2Cl_2$ is more complex than Eqn. 11 expresses explicitly. In this reaction, as will be shown below, the structure of the transition state itself depends sensitively on the solvent dielectric constant $\varepsilon$, so that both $\mu^2/r^3$ and the energy of the transition state species when transferred to vacuum also depend on $\varepsilon$. To account explicitly for these changes, an alternative formulation to Eqn. 11 is used:

$$\ln k = \ln A - \frac{1}{RT}\left\{E_a + \left(\Delta G_\varepsilon^{vac} - \Delta G_{\varepsilon_a}^{vac}\right) + \frac{N_A}{4\pi\varepsilon_o}\left(\frac{(\varepsilon_a-1)}{(2\varepsilon_a+1)}\left[\frac{\mu_{\ddagger,\varepsilon_a}^2}{r_{\ddagger,\varepsilon_a}^3} - \frac{\mu_A^2}{r_A^3} - \frac{\mu_B^2}{r_B^3}\right] - \frac{(\varepsilon-1)}{(2\varepsilon+1)}\left[\frac{\mu_{\ddagger,\varepsilon}^2}{r_{\ddagger,\varepsilon}^3} - \frac{\mu_A^2}{r_A^3} - \frac{\mu_B^2}{r_B^3}\right]\right)\right\}$$

$$(12)$$

In this expression, the terms $\Delta G_j^{vac}$ correspond to the energy for the transition state species (relative to the reactants) for the structure it has in a medium of dielectric constant *j when that structure is held fixed and the transition state complex is transferred to vacuum.* Similarly, the $\mu_{\ddagger,j}^2/r_{\ddagger,j}^3$ terms represent the properties of the transition state species when it sits in a medium of dielectric constant *j*. As in Eqn. 11, the effective activation energy represented by the quantity in curly brackets is divided into two contributions: a) the activation energy under ambient conditions, $E_a$, and b) remaining terms accounting for all factors arising from interaction with the solvent dielectric medium, collectively referred to as the correction factor. Note that the algebraic complexity in achieving this separation is a direct reflection of the physical complexity of a transition state which depends sensitively on its environment.



Calculation of the correction factor in either Eqn. 11 or 12 requires knowledge of a number of variables. The value of $\varepsilon$ for water has been measured experimentally as a function of temperature and pressure and can be determined from the correlation of Uematsu and Franck.[39] For stable molecules or reactants such as water or $CH_2Cl_2$, a number of sources exist that contain tabulated experimental values for the dipole moment, such as that by McClellan.[40] However, to determine the dipole moment and structure of the unstable transition state complex as well as the size of all relevant species requires *ab initio* quantum mechanical calculations, because no tabulated data exist for these quantities.

***Ab Initio* theory** *Ab initio* calculations have the great advantage of giving unambiguous results for a variety of fundamental molecular properties, thus limiting the uncertainties in final results and correlations. In addition to providing the necessary physical property values for use in calculating the rate constant correction factor, this approach provides valuable insight into the nature of reactions, such as environment-dependent solvent effects, as will be shown below.

Our *ab initio* calculations were carried out within the total energy plane wave density functional pseudopotential approach,[41] using the Perdew-Zunger parametrization[42] of the Ceperly-Alder exchange-correlation energy[43] and non-local pseudopotential of the Kleinmann-Bylander form[44] constructed using the optimization procedure of Rappe *et al.*[45] Our potential for carbon employed a Kleinmann-Bylander projector for the *s* channel, whereas for oxygen and chlorine we used a projector for the *p* channel. The electronic wave functions were expanded in a plane wave basis up to a kinetic energy cutoff of 40 Rydberg for a total of thirty-five thousand coefficients for each wave function. The calculations were carried out in a periodic supercell of dimensions 15 Å x 9 Å x 9 Å, allowing for sufficient separation between periodic images of the reacting species to minimize spurious effects, even for the rather elongated transition state. Note that all *ab initio* calculations described herein were performed in vacuum. The minimization of the energy over the



electronic coordinates was carried out using a parallel implementation of the conjugate-gradient technique of Teter *et al.*[46]

Ground state molecular structures were determined by moving the ionic cores along the Hellman-Feynman forces until all forces were less than 0.1 eV/Å. Candidate transition state structures were determined by holding the appropriate reaction coordinate (chosen as the distance between the carbon and oxygen atoms) fixed and relaxing the remaining coordinates along their respective forces until those forces were less than, again, 0.1 eV/Å. The vacuum energies of species ($G_i^{vac}$) are the direct output of this procedure. The dipole moment of each species was determined directly from the electronic charge density and nuclear arrangement, which are also direct byproducts of the minimization of the density-functional.

Determination of suitable "radii" $r_i$ for the different species for use in Eqns. 11 or 12 is more subtle. First, the species appearing in Eqns. 11 and 12 (the transition state in particular) are not spherical. To conform to Kirkwood's formulation in terms of spherical molecules, we use an average radius determined by the radius of a sphere equal in volume to the region about the molecule within which we expect the dielectric response from the solvent to be small, due to the exclusion of solvent molecules by short range repulsive forces. This radius we define as the equivalent spherical radius of the species.

Both the simplified equivalent spherical approach which we follow here and more sophisticated approaches require a procedure for estimating the region of diminished solvent dielectric activity resulting from the presence of the solute. To do this, we first generalize the concept of the van der Waals radius to a non-spherical surface, such that under experimental conditions it is very unlikely to find two species with their van der Waals surfaces overlapping. To determine the location of these surfaces, we performed *ab initio* calculations which show that the interaction energy between two water molecules rises rapidly above $k_bT$ when their oxygen nuclei



approach within 2.5 Å (0.25 nm) of each other.  The van der Waals surface for water thus extends 1.25 Å outward along this direction (which is in good agreement with the 1.4 Å radius determined from the experimental van der Waals equation of state for pure water[47]).  A similar analysis of the interaction between $CH_2Cl_2$ and $H_2O$ determined also that the van der Waals surface for $CH_2Cl_2$ extends 1.25 Å outward from the carbon nucleus in the direction directly opposite the leaving chlorine nuclei.  It would be computationally prohibitive to map out the van der Waals surfaces completely in all directions in this manner using *ab initio* calculations.  Instead, we determined this surface approximately as the constant electron density contour which passes through the two points identified.  This is reasonable because such a surface conforms to the shape of the molecule in much the same way as we expect the van der Waals surface would.

The van der Waals surface is relevant because the dielectric response of the solvent must drop to zero within the van der Waals surface of the solute.  This is because solvent molecules cannot contribute any charge to this region.  In general, the dielectric response of the solvent will vary continuously from zero up to its full value as we move outward from the van der Waals surface into the solvent over some distance characteristic of the size of the solvent molecules.  The region of depressed dielectric response, represented by the cavity in Kirkwood's formulation, thus extends outward beyond the van der Waals surface.  Simple model Monte Carlo calculations of isolated water molecules subjected to a constant electric field near a rigid interface showed this extra distance to be approximately 0.9 Å.  With the region of depressed dielectric response about $CH_2Cl_2$ thus determined to extend 2.15 Å from the carbon atom along the direction specified above, we approximated the volume of depressed dielectric response to be all points contained within the electron charge density contour passing through this point.  Using this same charge density contour value for all species, we determined all equivalent spherical radii needed in Eqn.12.



## VII.   RESULTS AND IMPLICATIONS OF *AB INITIO* CALCULATIONS

**Reactants**  Parameters describing the reactants $H_2O$ and $CH_2Cl_2$ provide an opportunity to validate our *ab initio* technique with existing measurements, and to check our choice of planewave cutoff and the construction of pseudopotentials in particular.  The structure of these molecules, as summarized in Table 3, is well reproduced.  For both, all bond lengths were within $\pm 0.03$ Å of their tabulated values.  The correct bent and tetrahedral geometries were observed for the stable equilibrium structure calculated for $H_2O$ and $CH_2Cl_2$.  The H-C-H and Cl-C-Cl bond angles for $CH_2Cl_2$ differed from their accepted values by only $0.5°$ and $0.2°$ respectively, while the H-O-H angle in the water molecule was within $1.6°$ of the established value of $104.5°$.

Figure 5 gives a three-dimensional representation of the full molecule showing a charge density contour containing 99.5% of the total charge density of the molecule.  From such densities the predicted dipole moment values are in good agreement with the experimental values (Table 3). The equivalent spherical radii for these species were also determined from the molecular charge densities according to the procedure described above.  The equivalent spherical radius for water as determined from the charge density contour chosen from the $CH_2Cl_2$ molecule is 2.14 Å.  The fact that this is very nearly the same as the sum of our 1.25 Å van der Walls radius for $H_2O$ and the 0.9 Å width of our dielectric buffer zone supports our notion of using a fixed charge density contour, and suggests that the appropriate value is reasonably transferable between species. Overall, our results confirm that our pseudopotentials are well suited for the atomic species involved in the reactions, and that our energy cutoff in the plane wave expansion is sufficient to describe the electronic wave functions of the system.

**Transition State**  With a validated *ab initio* approach, we can now proceed to look at the transition state.  Because of the demanding nature of the *ab initio* calculations, a complete search for all possible transition states would be prohibitive.  Fortunately, the chemistry of the present reaction is sufficiently well-understood that the pathway to search for the transition state is already



known. The $S_N2$ reaction mechanism occurs via a very specific stereochemistry, with the nucleophile attacking from the "back side"; the side directly opposite from the leaving group. To locate the transition state, we positioned a water molecule (as specified by the coordinates of its atomic nuclei) on the opposite side of the leaving Cl atom in a $CH_2Cl_2$ molecule, as depicted in Figure 6. The carbon atom was held fixed at the origin (*i.e.* the center of the supercell of the calculation) as the oxygen atom was forced to approach the carbon atom along the long axis of the supercell. At each C-O separation distance $\lambda$ describing the reaction coordinate for this reaction, the remaining degrees of freedom of the complex were relaxed without constraint, so that the molecules could change orientations at will until the equilibrium configuration and minimum energy were obtained. Rotational symmetry of the system ensures that this choice of reaction coordinate results in no loss of generality.

To ensure that the molecules did not lock into false local minima during the approach, we explored the two most likely pathways for $H_2O$-$CH_2Cl_2$ approach. In the first, (a) the two hydrogen atoms from $H_2O$ began in the same plane as the carbon and leaving chlorine atoms. This approach minimizes the energy by positioning the hydrogen atoms from the $H_2O$ as far as possible from the two hydrogen atoms from $CH_2Cl_2$ (see Figure 6). In the second approach, (b) the $H_2O$ molecule was rotated 90° so that its hydrogen atoms now began in the plane orthogonal to that containing the $H_2O$ molecule in (a). In this alignment, one proton from the $H_2O$ molecule approaches the non-leaving Cl and the other approaches the two hydrogen atoms of the $CH_2Cl_2$ molecule. For the case explored, at a typical value of $\lambda$ where the transition state complex was found to occur, the results from orientation (b) showed a significantly greater energy than from orientation (a) by about 20 kJ/mol. We thus focused our effort on pathway (a).

In our cell of length 15 Å, the largest possible value of the generalized coordinate, $\lambda = 7.5$ Å, corresponds to two separate, stable $H_2O$ and $CH_2Cl_2$ molecules. We confirmed that the resulting equilibrium structures at $\lambda = 7.5$ Å match our results for the two isolated species. As $\lambda$



decreases, the two reactants begin to feel the effects of their proximity. The most critical region occurs over the range 2.0 Å < $\lambda$ < 1.4 Å. As $\lambda$ decreases through this range, the dipole moment $\mu$ increases by more than a factor of two (Figure 7), and the total energy (in vacuum) of the system relative to the isolated reactants at $\lambda$ = 7.5 Å increases by almost a factor of three ($\varepsilon$ = 1 curve in Figure 8). This dramatic variation in the dipole moment over such a small range of the reaction coordinate has profound consequences for the reaction as the nature of its dielectric environment changes, as we shall discuss below. However, because there is not much change in the physical size of the transition state over such a limited range of reaction coordinate, variations in the volume of the activated complex are minimal (*e.g.* $\pm$ 0.1 Å in the equivalent spherical radius over the same range in $\lambda$.).

Although the total energy of the $H_2O$ and $CH_2Cl_2$ pair clearly changes as the two molecules are brought closer together, in vacuum it does not show the usual profile one would expect for two reacting species. Note that the reaction profile, or the energy relative to isolated reactants as a function of the reaction coordinate, increases monotonically with decreasing separation when $\varepsilon$ = 1 (Figure 8). This implies that there is always repulsion and never attraction between the species and thus no stable reaction product under vacuum conditions.

To represent the behavior of $H_2O$ and $CH_2Cl_2$ in a solvent of dielectric constant $\varepsilon$, the energy of solvation must be added to the vacuum energy. Using the relation for $\Delta G_{solv, i}$ derived from Kirkwood theory (Eqn. 6) along with the appropriate *ab initio* values of $\mu(\lambda)$ and $r(\lambda)$ already correlated, the total energy for the $H_2O/CH_2Cl_2$ pair can be calculated in a solvent of arbitrary dielectric constant $\varepsilon$ at any separation distance $\lambda$. The results (Figure 8) show a dramatically different behavior for $H_2O$ and $CH_2Cl_2$ when surrounded by a water solvent. Figure 8 also shows the reaction profiles for a variety of different dielectric strengths $\varepsilon$ (which at fixed pressure is a function only of temperature). For values of $\varepsilon$ > 3.6, the reaction profiles exhibit the local maximum and minimum characteristic of the formation of a product. Because the *ab initio*



technique used results in the minimum configuration energy at any position $\lambda$, the profiles in Figure 8 represent the lowest energy pathway for reaction at any given $\varepsilon$. Along any profile, therefore, we identify the value of $\lambda$ where the local maximum in energy occurs as the location of the transition state complex. The location of the local minimum likewise corresponds to the product. This maximum and minimum in the profiles become more pronounced as $\varepsilon$ increases. Thus, the influence of the solvent on the $CH_2Cl_2$-$H_2O$ transition state, and therefore on the nature of the reaction itself, is substantial.

Close examination of the reaction profiles in Figure 8 reveals that the local maximum energy points do not all occur at the same value of $\lambda$, but happen at slightly smaller values of $\lambda$ as $\varepsilon$ decreases. Thus for every $\varepsilon$, there is a particular, slightly different, value of $\lambda$ at which the transition state occurs. As a result, the structure and physical properties of the transition state, already functions of $\lambda$, are ultimately functions of $\varepsilon$ (or $T$ at fixed pressure, from the correlation of Uematsu and Franck[39]), and can be traced and correlated as such.

Over the full range of values for $\varepsilon$ appearing in our experiments, the structural transformations predicted along the reaction pathway are consistent with the known stereochemistry of the $S_N2$ reaction and with empirical rules for intermolecular transition states. As $\varepsilon$ varies from 80 (22°C) to 1.3 (600°C), our predicted location of the transition state occurs at values of $\lambda$ of 1.72 Å to 1.63 Å respectively (Figure 9). Values of the other predicted bond lengths for the transition state corresponding to both limiting values of $\lambda$ are displayed in Table 4, along with the accepted bond lengths for stable molecules. The calculations show that as the $H_2O$ molecule approaches the carbon atom of the $CH_2Cl_2$ molecule (*i.e.* $\lambda$, or C-O bond length, decreases), the Cl atom opposite the approaching $H_2O$ begins to move further away from the carbon atom (*i.e.* C-Cl bond length increases), as expected for this reaction. Note that the C-O bond length calculated for the transition state is about 0.2 - 0.3 Å greater than the normal C-O bond length of 1.41 Å. Likewise, the C-Cl bond length (for the leaving Cl) calculated for the transition



state is about 0.5 - 0.7 Å greater than the normal C-Cl bond length of 1.77 Å.  These variations in bond length are consistent with the empirical rule cited by Benson[49] that bonds in the process of being formed or broken in intermolecular transition states for molecule-molecule reactions should be 0.3 - 0.5 Å greater than their nominal length in stable molecules.

Although the lengths of bonds not participating in the reaction do not deviate far from their values in the isolated reactants (Table 4), the orientations of these bonds change significantly along the reaction pathway.  Specifically, the two C-H bonds and the C-Cl bond with the non-leaving Cl fold over from their tetrahedral geometry to a planar arrangement perpendicular to the direction of the incoming water molecule, as shown in Figure 6.  This planar configuration for the transition state corresponds precisely to the stereochemistry expected for an $S_N2$ reaction.

Although the reaction coordinate $\lambda$ at which the transition state occurs varies by only 0.1 Å over the range of $\varepsilon$ sampled in our experiments, the calculated value of the dipole moment for the transition state varies by almost 2 Debye units (or about 25%) (Table 4).  This is because the value of $\mu$ for the $CH_2Cl_2$-$H_2O$ pair is particularly sensitive to the value of $\lambda$ in the region where the transition state was found to occur (*i.e.*, the steep portion of the curve in Figure 7).  One should also note that the magnitude of $\mu$ for the transition state in this active region is predicted to be more than four times larger than that for either of the reactants, which again is consistent with expectations for an $S_N2$-type reaction between neutral reactants resulting in a charged leaving group.  Both trends observed for $\mu$ will be shown to have a profound effect on the hydrolysis rate. In contrast, the equivalent spherical radius varies by only 0.02 Å over the relevant range of $\lambda$ and therefore is not the determining factor in the behavior of the rate constant.

**Implications of Solvent Effects on Reaction Rate**  Figure 8 exhibits three major trends in the behavior of the transition state as the dielectric nature of the solvent varies.  The detailed



information which the *ab initio* calculations provide allow us to understand these trends in terms of the behavior of the reacting molecular complex, summarized as follows:

(1) As already mentioned earlier, the transition state occurs at smaller values of $\lambda$ as $\varepsilon$ decreases. The reason for this is that as $\varepsilon$ decreases, the dielectric solvent environment becomes less effective in its role in stabilizing the reactants. To compensate, the transition state moves to less negative values of $\lambda$ (smaller intermolecular distances), where the dipole moment of the complex is stronger.

(2) As $\varepsilon$ decreases, the energy associated with the transition state (local maximum) increases. Since the difference in energy between that of the transition state and that of the two reactants (which is what is plotted in Figure 8) is defined as the activation energy for the reaction, the calculations predict an increasing activation energy with decreasing solvent polarity. This increase in energy can be traced to the fact that despite the increase in $\mu$ described in item 1, the vacuum energy of the transition state complex (Figure 8) rises so quickly as $\lambda$ decreases that the net effect is for energy to increase with decreasing $\varepsilon$. The resulting trend is consistent with a decreasing reaction rate, as observed experimentally.

(3) The local maximum and minimum in energy become less pronounced and move closer together (*i.e.* occur at the same value of $\lambda$) as $\varepsilon$ decreases and vacuum conditions are approached. By the point where $\varepsilon = 3.6$ (corresponding to 386°C), the maximum and minimum become indistinguishable and are replaced by a small plateau (inflection point). This decreasing energetic separation between the maximum and minimum points as $\varepsilon$ decreases indicates a reaction whose product is becoming less stable and more likely to recross the transition state barrier and return to the original reactants. This too would lead to a reduced net rate of reaction.



Item (2) above provides the theoretical explanation for the experimental observation of slower hydrolysis kinetics of $CH_2Cl_2$ upon heating from sub-to supercritical conditions. Table 5 lists values of the activation energy for $CH_2Cl_2$ hydrolysis calculated from our *ab initio* results at various temperatures and corresponding values of $\varepsilon$. A value of 130 kJ/mol is predicted for the ambient activation energy $E_a$ at 25°C. The accuracy of the local density approximation used in our calculations for an absolute reaction barrier is generally to be considered to be on the order of 10 to 20 kJ/mol because of the large difference in the variation in the molecular electron density in going from the isolated reactants to the transition state. Our determination of the reaction barrier in ambient water also depends upon several approximations inherent in the Kirkwood approach. More detailed calculations[50] place the error associated with these approximations also at around 10 kJ/mole when comparing isolated reactants and the transition state. Given these uncertainties, our *ab initio* result for $E_a$ compares quite well with the value of 107 kJ/mol found experimentally by Fells and Moelwyn-Hughes[16] at 115°C.

Finally, item (3) raises the issue as to the nature of the reaction when $\varepsilon < 3.6$ (*i.e.* above 386°C at 246 bar). Within simple transition state theory, the disappearance of the maxima and minima in the reaction profile that our calculations have revealed for $T \geq 386$°C would imply that there is no stable product and that the reaction comes to a halt. However, because we have not explored the entire possible phase space of this two-step reaction, it would be precipitous to construct a model where the reaction rate is taken to be strictly zero above this temperature. Rather than this, which is equivalent to setting the reaction barrier to infinity above $T = 386$°C, we have taken a more conservative approach. We allow for the possibility for further reaction above this temperature by extrapolating the value of the energy barrier as a continuous function of temperature. The particular form we chose for extrapolation was to take the energy barrier to be the value of the energy at the inflection point in the profiles above $T = 386$°C, which connects continuously to the maxima and minima at this temperature. Our results show that the reaction rate



has slowed enough by 386°C that even with this more conservative extrapolation, not much conversion occurs above this temperature (see below).

**Generation of the Kirkwood Rate Constant Correction Factor**  With the critical parameters of the $CH_2Cl_2$-$H_2O$ transition state and the isolated reactants (*i.e.* vacuum energy, dipole moment, and equivalent spherical radius) now determined, we can apply Eqn. 12 to compute the reaction rate constant.  We separate out and formally define the *Kirkwood correction factor to the rate constant*, $\Phi$, relative to ambient conditions ($\varepsilon_a$), so that

$$\ln k = \ln A - \frac{E_a}{RT} + \Phi \quad , \tag{13}$$

where

$$\Phi = -\frac{1}{RT}\left(\left(\Delta G_\varepsilon^{vac} - \Delta G_{\varepsilon_a}^{vac}\right) + \frac{N_A}{4\pi\varepsilon_o}\left(\frac{(\varepsilon_a-1)}{(2\varepsilon_a+1)}\left[\frac{\mu_{\ddagger,\varepsilon_a}^2}{r_{\ddagger,\varepsilon_a}^3}-\frac{\mu_A^2}{r_A^3}-\frac{\mu_B^2}{r_B^3}\right]-\frac{(\varepsilon-1)}{(2\varepsilon+1)}\left[\frac{\mu_{\ddagger,\varepsilon}^2}{r_{\ddagger,\varepsilon}^3}-\frac{\mu_A^2}{r_A^3}-\frac{\mu_B^2}{r_B^3}\right]\right)\right)$$

$$\tag{14}$$

Here, the reactants A and B represent $CH_2Cl_2$ and $H_2O$.  Breaking up the expression for the reaction rate constant in this way into an Arrhenius form and a correction term increases the reliability of our prediction by collecting into a single term the major sources of error involved in the local density approximation employed in our *ab initio* calculations and the simple dipole - sphere model from Kirkwood theory.  The local density approximation reproduces the energy differences among configurations with similar electron densities (such as two transition states at slightly different $\lambda$) much better than between configurations with far different densities (such as the transition state and the isolated reactants).  Preliminary results indicate similar trends for the approximations in Kirkwood theory.[50]  Thus, while the overall reaction barriers ($E_a$ - $RT\Phi$) are reliable to only 20 kJ/mol, the differences among barriers for different $\lambda$ (contained in the $\Phi$ term) are much more reliable.  Most of the uncertainty is contained in $E_a$, which may be determined from experimental data.  Note also that $\Phi$ contains no fitted or adjustable parameters.



Table 6 and Figure 10 show $\Phi$ as a function of the dielectric constant and temperature. These calculated results indicate that $\Phi \le 0$ over the entire temperature range of interest (25-600°C), consistent with the expected tendency to retard the reaction as the water solvent becomes less polar. In the subcritical region where water is still quite polar, $\Phi$ is small; however, in the range of temperatures just beyond the critical point where $\varepsilon$ drops by an order of magnitude, $\Phi$ increases dramatically. By 550°C, well into the supercritical regime, we predict the Kirkwood correction factor to decrease the rate constant by 7.5 natural log units (three orders of magnitude) relative to that predicted by simple Arrhenius behavior. Because the density of the solvent drops by only a single order of magnitude over the same range of conditions, our results demonstrate that for this reaction, the rate is much more strongly correlated to the dielectric constant than to the density. To facilitate calculations, four separate polynomial functions were fit to represent $\Phi(T)$ in Figure 10 for the temperature ranges 25-374°C, 374-386°C, 386-525°C, and 525-600°C (see Table 7).

To illustrate the effect that the Kirkwood correction factor $\Phi$ can have on $CH_2Cl_2$ hydrolysis kinetics, we have used it to modify the empirical correlation for the rate constant, $k_{FMH}$, proposed by Fells and Moelwyn-Hughes[16] (see Eqn. 1):

$$\ln k' = \ln k_{FMH} + \Phi \qquad (15)$$

Figure 11 compares the value of $\ln k_{FMH}$ to its corrected value $\ln k'$ over the much wider temperature range of 25 to 600°C (compared to the 80-150°C range originally studied by Fells and Moelwyn-Hughes). Note that in the limited temperature range of Fells and Moelwyn-Hughes' experimental data, both $k'$ and $k_{FMH}$ are essentially the same, verifying that solvent effects are negligible in this regime. As temperature increases, the two rate constant values begin to diverge from each other due to the increasing magnitude of $\Phi$. In the supercritical region at $T > 400°C$, $k'$ is substantially lowered to a value comparable to that seen around 165°C, where the rate is relatively slow. Solvent effects decrease the uncorrected hydrolysis rate by more than a factor of 1700 at 550°C (Table 6). Thus inclusion of the Kirkwood correction factor in the form of the rate



constant yields rate behavior consistent with our experimental observation of a slowing reaction with increasing temperature.

## VIII.   DETERMINATION OF HYDROLYSIS GLOBAL KINETIC RATE EXPRESSION

As mentioned earlier, the neutral/acidic hydrolysis reaction of $CH_2Cl_2$ occurs in a two stage process.  Because the substitution reaction is the slow step in the sequence, the overall rate of reaction incorporating the Kirkwood correction factor $\Phi$ is:

$$R_{hyd} = \frac{d[CH_2Cl_2]}{dt} = -\left(Ae^{(-(E_a/RT)+\Phi)}\right)[CH_2Cl_2]^a[H_2O]^b \qquad (16)$$

where $A$ and $E_a$ represent the pre-exponential factor and activation energy under ambient conditions, and $a$ and $b$ are the reaction orders with respect to $CH_2Cl_2$ and $H_2O$, respectively.  The explicit dependence on $H_2O$ concentration reflects the order of magnitude variation of solvent density (over the subcritical section of the preheater) in our experiments.  As a result, despite being always in excess relative to $CH_2Cl_2$, the $H_2O$ concentration was not constant, and therefore cannot be lumped *a priori* into the rate constant.

Optimal values of the kinetic parameters $A$, $E_a$, $a$ and $b$ were determined by nonlinear regression of our experimental data.  This allows us to combine the results of the *ab initio* calculations describing the strong variations in the energy and dipole moment of the transition state which are embodied in $\Phi$, with the experimental results, to yield a quantitatively reliable rate expression.  The unknown parameters were optimized to match as closely as possible the observed $CH_2Cl_2$ conversions.  The regression procedure used is a two-step process.  First, for a given choice of $A$, $E_a$, $a$ and $b$ (regarded as the independent parameters), one computes the total $CH_2Cl_2$ conversion (the dependent parameter) at the end of the preheater and main reactor system using the same differential heat and mass balance equations for temperature and $CH_2Cl_2$ concentration described earlier (Eqns. 2, 3, and 5).  In this case, however, Eqn. 16 was used for the hydrolysis



rate expression $R_{hyd}$ instead of Eqn. 4 or 15. Second, one adjusts the independent parameters iteratively until a best-fit to the experimental conversion results is obtained. The latter process was performed using the multivariable Powell SSQMIN algorithm[51] as the nonlinear optimization routine.

The full four-parameter regression to the hydrolysis experimental data showed the best-fit values of $b$, $A$ and $E_a$ to be highly correlated, making it difficult to extract meaningful values. (Similar behavior was noticed by Holgate *et al.*[21] in performing a four-parameter regression for the water gas shift reaction between CO and $H_2O$.) As the expected reaction order for $H_2O$ (given the nature of the $S_N2$ reaction) is first order, $b$ was set to unity in all further regression attempts.

With $b = 1$, the three-parameter regression for $A$, $E_a$, and $a$ yielded the following global rate expression for $CH_2Cl_2$ hydrolysis:

$$R_{hyd} = \frac{d[CH_2Cl_2]}{dt} = -\left(10^{16.5\pm4.6}e^{(-(210\pm40/RT)+\Phi)}\right)[CH_2Cl_2]^{1.52\pm0.67}[H_2O] \tag{17}$$

where $R_{hyd}$ is in units of mol/(L s), activation energy is measured in kJ/mol, and the concentrations are in mol/L. All parameter uncertainties are quoted at the 95% confidence level, and $\Phi$ is as given by Eqn. 14, with values determined through the polynomial correlations given in Table 7. The regressed value of $210\pm40$ kJ/mol for the ambient activation energy ($E_a$) includes all experimental uncertainties, modeling uncertainties, and variability due to the number of parameters in the regression. Therefore this value is not unreasonable when compared to the parameter-free Kirkwood-*ab initio* determined value of 130 kJ/mol (Table 5).

Figure 12 compares total $CH_2Cl_2$ conversion calculated with the rate expression of Eqn. 17 as a function of experimental residence time and sandbath or final reactor temperature with the experimental hydrolysis data and the results of the simple first-order model of Fells and Moelwyn-



Hughes.[16]  Predicted values of conversion at both the end of the preheater and after the main

reactor are presented.  Figure 12 demonstrates that accounting for solvent dielectric and density

effects improves significantly the prediction of conversion both qualitatively and quantitatively over

simply extrapolating the experimental Arrhenius correlation of Fells and Moelwyn-Hughes.  Use

of the Kirkwood correction factor $\Phi$ in Eqn. 17 also leads to a much reduced further conversion of

$CH_2Cl_2$ in the supercritical main reactor itself, in agreement with our experimental observations.

For sandbath temperatures below 525°C (total residence times < 16.5 s), predicted conversions at

the end of the preheater and main reactor were within the experimental error of the data, and were

essentially identical at 450°C (14 s), exactly as observed.  The ability of Eqn. 17 to capture this

behavior, not predicted by extrapolation of the single Arrhenius rate expression of the Fells and

Moelwyn-Hughes model, underscores the importance of the Kirkwood correction factor on the

kinetics of polar reactions occurring in solvent media undergoing appreciable changes in dielectric

constant.  At higher temperatures and longer residence times, the fit of predicted conversion at the

end of the reactor from Eqn. 17 still matches the data well, but is increasingly higher than that

predicted at the end of the preheater.  Although the maximum difference between predicted

preheater and main reactor conversions was only 16%, it is higher than the 3% difference observed

experimentally at a sandbath temperature of 575°C (20 s) (Table 2).

Figure 13 shows how predicted $CH_2Cl_2$ conversion (from Eqn. 17) varies as a function of

temperature within the preheater tubing.  The graph presents results for four runs, which include

the experimental sandbath temperature and residence time extremes.  In all cases, most of the

conversion occurs below 390°C, with a smaller amount occurring under supercritical conditions.

In the temperature range from  390 - 540°C (well above the critical point), the model predicts no

noticeable conversion.  The small increase in conversion that appears after about 540°C for the

higher sandbath temperature runs, which is correlated to the somewhat higher main reactor

conversions under these conditions shown in Figure 12, appears to be a feature of the model at

these highest temperatures.  It arises from our conservative treatment of the reaction rate as



continuing after the disappearance of the energy maxima and minima in the reaction profile above 386°C, as discussed above.

The three-parameter regression of Eqn. 17 contains the theoretical value of $a = 1$ within experimental uncertainty. Accordingly, the regression was rerun with $a$ and $b$ fixed at unity. This two parameter regression yielded the following second-order overall rate expression:

$$R_{hyd} = \frac{d[\text{CH}_2\text{Cl}_2]}{dt} = -\left(10^{12.9 \pm 1.0} e^{(-(180 \pm 14/RT) + \Phi)}\right)[\text{CH}_2\text{Cl}_2][\text{H}_2\text{O}] \qquad (18)$$

where units and confidence levels for uncertainties are the same as for Eqn. 17. In Figure 14, calculated values of total $\text{CH}_2\text{Cl}_2$ conversion using Eqn. 18 are compared to the experimental data and Fells and Moelwyn-Hughes model in a similar plot as that of Figure 12. The results are similar to those seen in Figure 12, with the fit of predicted conversion at the end of the reactor to the experimental data being equally as good (similar $\chi^2$), as expected. The only exception is that the increasing difference between preheater and reactor predicted conversions at higher sandbath temperatures/longer residence times is greater than that generated from use of Eqn. 17 in Figure 12. The ambient activation energy of 180 kJ/mol is, however, in somewhat better agreement with the Kirkwood-*ab initio* value of 130 kJ/mol.

## IX.  DISCUSSION  AND  CRITIQUE

Given the simplistic nature of the Kirkwood model and its many assumptions, the correction factor $\Phi$ captures the essential form and magnitude of the diminishing effects of the solvent on the $\text{CH}_2\text{Cl}_2$ hydrolysis rate remarkably well. Our results establish that the dielectric coupling between the reacting species and the solvent is the major factor leading to the non-Arrhenius behavior of the reaction rate under supercritical conditions.



The modest discrepancies between our prediction of preheater and main reactor conversions at higher sandbath temperatures, however, suggests that our physical model is still incomplete. There are three possible explanations for this behavior: (1) more than one reaction mechanism occurred over the extended temperature range covered in the experiments, confounding our attempts to model the conversion as proceeding by a single reaction; (2) our conservative continuation of the reaction rate beyond the disappearance of the local energy maximum along the reaction profile for $\varepsilon < 3.6$ ($T > 386°C$) is unrealistic as the reaction rate might become negligible beyond this temperature; (3) our picture is essentially correct and the discrepancies which we observe are purely due to quantitative errors associated with the simplifying assumptions in the Kirkwood model itself.

A more detailed view of the first possibility is that the reaction may be proceeding via different pathways at different temperatures (*i.e.* subcritical polar vs. supercritical free radical pathways), or that one or more additional reactions involving $CH_2Cl_2$ become important at higher temperatures. However, there is no direct evidence either in the literature or the experimental product distribution from our studies that suggests that any significant secondary reactions could have occurred.[5] In addition, recent isothermal batch experiments on $CH_2Cl_2$ hydrolysis in the presence of Hastelloy C-276 beads confirm no effect of the metal on hydrolysis kinetics[52].

With the second possibility, the precise nature of the change in the reaction signaled by the disappearance of the local energy maximum and minimum at $T = 386°C$ ($\varepsilon = 3.6$) is uncertain. Without further exploration of the reaction phase space, we chose the more conservative route of allowing for a finite reaction rate above this temperature. Even with this approach of extrapolating the decreasing reaction rate near $T = 386°C$, most of the $CH_2Cl_2$ hydrolysis is still predicted to occur in the subcritical region. It does appear, however, that the magnitude of the correction factor at higher temperatures (*i.e.* > 540°C) is not large enough to counteract the increasing magnitude of the Arrhenius component of the rate expression. The comparison of this with the experimental



results may signify that the disappearance of the maxima and minima should be taken to mean that the reaction stops completely above 386°C. One interesting direction for future work is therefore to explore the hydrolysis reaction phase space more fully.

With regards to the third possibility, shortcomings of the Kirkwood model are readily apparent. First are the structural idealizations: (a) molecules are not spherical, so that one cannot properly characterize the size of a molecule like $CH_2Cl_2$ or its hydrolysis transition state complex by a single radius; and (b) the charge distribution of a molecular complex is distributed in space and not concentrated at a point so that in the vicinity of such a complex, where the coupling to the solvent is the strongest, the charge distribution cannot be captured by the first few terms in a multipolar expansion. Second, is the manner in which the dielectric response of the solvent is treated on the molecular level: (c) by using a local dielectric function $\varepsilon$ outside of the molecule, Kirkwood theory ignores molecular correlations which are needed to describe such effects as hydrogen bonding; and (d) the theory also assumes an unrealistically sharp transition of the solvent dielectric as one approaches the interior of the molecule. Minor flaws include an uncertainty as to how to define the interior and exterior regions of the molecule and the value of the dielectric constant which one should use in the interior solute region. Our *ab initio* work described above, however, mitigates the uncertainties as to the boundary of the molecule, and it is clear that in self-consistent *ab initio* calculations, the dielectric constant inside of the molecule should be taken to be that of vacuum.

Both continuum and molecular-level studies are underway to evaluate the importance of the above simplifications and to find ways to correct for these deficiencies. Preliminary results show that the impact of using a more realistic shape for the molecular cavity and the complete *ab initio* charge distribution as opposed to a spherical cavity containing only the dipole moment of the distribution is relatively mild, changing the correction factor by less than 10%.[50] Thus, while future improved results may refine some of the quantitative details of our regression, they will



preserve the overall picture. Finally, molecular dynamics (MD) simulations of large numbers of water molecules interacting with the potential field of the solute as determined by *ab initio* are underway to evaluate the impact of more realistic microscopic treatment of the dielectric response of the solvent.

## X.  SUMMARY AND CONCLUSIONS

Experimental evidence shows that hydrolysis is a significant pathway of destruction for $CH_2Cl_2$, with most breakdown occurring at subcritical temperatures. Slower hydrolysis rates with little additional $CH_2Cl_2$ conversion occurred under supercritical conditions. With these complexities, a simple, Arrhenius form for the hydrolysis rate constant does not match the experimentally observed trends over the extended temperature range investigated.

There are qualitative reasons to believe that the $CH_2Cl_2$ hydrolysis reaction should slow down as $T$ increases and $\varepsilon$ decreases. The changing nature of the water solvent when heated from sub- to supercritical conditions decreases its ability to stabilize the polar transition state complex. Kirkwood Theory provides an effective framework and *ab initio* techniques provide accurate microscopic parameters with which to quantify these solvent effects.

The *ab initio* predicted stereochemistry and physical properties of both reactants and their transition state are consistent with expectations. Increases in the activation energy and a changing reaction profile with decreasing $\varepsilon$ (increasing $T$) provide a mechanism for reducing the reaction rate at higher temperatures. The hydrolysis rate is predicted to decrease by about three orders of magnitude at 550°C relative to that predicted without accounting for these solvent-related effects. The presence of the screening provided by the solvent is necessary to stabilize the transition state. Thus, the transition state and activation energy for the hydrolysis reaction are highly sensitive to the changing dielectric constant as water becomes supercritical. The magnitude of these effects are properly captured and correlated by a Kirkwood correction factor to the rate constant, $\Phi$ (Eqn. 14).



A regressed global rate expression for hydrolysis incorporating $\Phi$ (Eqn. 17, 18) was developed to predict the trends in the experimental data and provide a model of the reaction rate over an extended temperature range of interest in engineering applications.

## APPENDIX

**Determination of Overall Heat Transfer Coefficient Values**  Values of the overall heat transfer coefficient $U_i$ were needed for use in Eqn. 2 in order to calculate the temperature profile in the preheater tubing of the experimental system.  For these calculations, $U_i$ was considered to be a function of $z$ and was computed at each step in the numerical integration using the usual sum of resistances formula:

$$\frac{1}{U_i} = \frac{1}{h_i} + \frac{t_w d_i}{k_w d_{LM}} + \frac{d_i}{d_o h_o} \qquad\qquad \text{(A-1)}$$

where $h_i$ and $h_o$ are the internal and external heat transfer coefficients, respectively; $d_i$, $d_o$, and $d_{LM}$ are the inner, outer, and log mean diameter, respectively; $t_w$ is the wall thickness; and $k_w$ is the wall thermal conductivity.

Calculation of $h_i$ in this system required consideration of a number of important factors and phenomena including substantial changes in physical properties (particularly near the critical point), geometric and flow effects, and the proper coupling of forced and natural convection.  A detailed analysis is provided by Marrone.[52]  Six correlations were taken from the literature to cover the three different spatial orientations of sections of the tubing (vertical downflow, horizontal flow, and vertical upflow) and whether the flow was laminar or turbulent.  Sources of these correlations are listed in Table A1.  All of these correlations are given in terms of the Nusselt number $Nu$ and can generally be written as follows:

$$Nu = \frac{h_i d_i}{k} = f\left(Re, Pr, Gr, \frac{\rho_w}{\rho}, \frac{\overline{C}_p}{C_p}\right) \qquad\qquad \text{(A-2)}$$



where $Re$ is the Reynolds number, $Pr$ is the Prandtl number, $Gr$ is the Grashof number, $\rho_w$ and $\rho$ are the density at the wall and in the bulk respectively, and $\overline{C}_p$ and $C_p$ are the average heat capacity between wall and bulk temperatures and the bulk heat capacity respectively.

Values for the tube wall conductivity, $k_w$, were calculated from a temperature-dependent empirical equation fit to thermal conductivity data for Hastelloy C-276.[57] The external heat transfer coefficient, $h_o$, was assumed constant over the tubing length for a given sandbath temperature. Values for $h_o$ were chosen for each experiment so that the model predicted the measured value of the mixing tee temperature at the end of the preheater tubing. For the entire range of operating conditions, the necessary values of $h_o$ ranged from 69 to 316 W/m$^2$K, varying by less than a factor of 5. Uncertainty in the measured mixing tee temperature had only a modest effect on calculated $h_o$ values; a $\pm$ 1°C change in the mixing tee temperature resulted in only a 3% change in $h_o$. The assumption that $h_o$ was constant is reasonable, given the uniform thermal conditions that existed in the fluidized sandbath in which the preheater tubing and main reactor were immersed. The experimentally fitted values of $h_o$ agree reasonably well with values of about 260-450 W/m$^2$K calculated from the correlation of Vreedenberg[58] for a horizontal tube in a fluidized sandbath.

Based on the values of $h_i$, $k_w$, and $h_o$ calculated for the conditions of our experiments, most of the heat transfer resistance occurred external to the preheater tubing. Typically, $h_i$ was about an order of magnitude larger than $h_o$. The resulting temperature-time profiles indicate preheater residence times of 7 to 17 s, with about 70-80% of the total residence time of the fluid spent below 390°C, where water is still sufficiently polar for $CH_2Cl_2$ hydrolysis to occur. Temperature predictions also agree with independent corrosion measurements in that a range of 110-350°C was predicted for the region of the preheater tube wall where most of the observed corrosion occurred. This hot, but still subcritical, temperature range is where the tube wall should be most susceptible to chloride-induced corrosion.



## ACKNOWLEDGMENTS

The authors would like to thank the U.S. Army Research Office (grants DAAL03-92-G-0177, DAAH04-93-G-0361, DAAG04-94-G-0145, and DAAH04-96-1-0174) under the University Research Initiative program supervised by Dr. Robert Shaw, and the National Institute of Environmental Health Sciences (grant No. 5 P42 ESO4675-10), for their partial support of this research. TAA is supported in part by a grant from the Alfred P. Sloan Foundation (BR-3456). Computational support was provided by the MIT Xolas prototype SUN cluster. Prof. Phil Gschwend of the Department of Civil and Environmental Engineering at MIT is gratefully acknowledged for providing insight on the nature of the subcritical $CH_2Cl_2$ hydrolysis reaction.

# List of Tables



## List of Figures





| | | |
|---|---|---|
| **Figure 3** | Experimental and Predicted $CH_2Cl_2$ Hydrolysis Conversion As a Function of Sandbath Temperature and Residence Time. Total pressure = 246 bar; error bars indicate 95% confidence intervals. Model values generated using the first order rate constant correlation of Fells and Moelwyn-Hughes (Eqn. 1). Note that residence time does not scale linearly with sandbath temperature. | |
| **Figure 4** | The Kirkwood Model for a Solute Molecule in Solution. | |
| **Figure 5** | Three-Dimensional Representation of $CH_2Cl_2$ Molecule in Terms of Charge Density Determined from *Ab Initio* Calculations. Boundary of molecule shown includes 99.5% of charge density. Center spheres indicate locations of atomic nuclei. | |
| **Figure 6** | Stereochemistry of Species in $CH_2Cl_2$ Hydrolysis Determined from *Ab Initio* Calculations. (a) reactants (b) transition state complex. | |
| **Figure 7** | Plot of Dipole Moment $\mu$ of the $H_2O$ - $CH_2Cl_2$ Pair as a Function of the Carbon - Oxygen Separation Distance $\lambda$. All values of $\mu$ were determined from charge densities calculated via ab initio simulation. | |
| **Figure 8** | Energy of the $H_2O$ - $CH_2Cl_2$ Pair in Solution as a Function of Their Carbon - Oxygen Atom Separation Distance $\lambda$ and dielectric constant $\varepsilon$. All energy values are determined from *ab initio* calculations and Eqn. 6, and are referenced to the sum of energies for the individual $H_2O$ and $CH_2Cl_2$ molecules. | |
| **Figure 9** | Graph of the Carbon-Oxygen Separation Distance Corresponding to the $CH_2Cl_2$ Hydrolysis Transition State ($\lambda_{TS}$) as a Function of Temperature. | |
| **Figure 10** | Kirkwood Correction Factor $\Phi$ and Water Dielectric Constant $\varepsilon$ as a Function of Temperature. The curve for $\Phi$ is determined from *ab initio* data in Figures 7 and 8 and Table 4. Polynomial correlations for this curve are given in Table 7. Values of $\varepsilon$ determined from the correlation of Uematsu and Franck[39] at 246 bar. | |
| **Figure 11** | Plot of the Natural Logarithm of the Fells and Moelwyn-Hughes Rate Constant with and without the Kirkwood Correction Factor ($k'$ and $k_{FMH}$ respectively) as a Function of Temperature. Values of $k'$ and $k_{FMH}$ are calculated as defined in Eqns. 15 and 1. The temperature range of the Fells and Moelwyn-Hughes[16] experiments on neutral $CH_2Cl_2$ hydrolysis (80-150°C) is indicated. | |
| **Figure 12** | Experimental and Predicted $CH_2Cl_2$ Hydrolysis Conversion as a Function of Sandbath Temperature and Residence Time (3 parameter model). FMH predicted conversion is from the first order model of Fells and Moelwyn-Hughes[16] with the rate constant as given in Eqn. 1. MIT + Kirkwood predicted conversion uses the three parameter global rate expression of Eqn. 17 developed in the present work. | |
| **Figure 13** | Predicted $CH_2Cl_2$ Conversion Versus Temperature in the Preheater Tubing for Four Different Sandbath Temperatures and Flow Rates. All values calculated using the global rate expression of Eqn. 17 | |
| **Figure 14** | Experimental and Predicted $CH_2Cl_2$ Hydrolysis Conversion as a Function of Sandbath Temperature and Residence Time (2 parameter model). FMH predicted conversion is from the first order model of Fells and Moelwyn-Hughes[16] with the | |



rate constant as given in Eqn. 1. MIT + Kirkwood predicted conversion uses the two parameter global rate expression of Eqn. 18 developed in the present work for *a* and *b* forced to 1.



# Figure 1

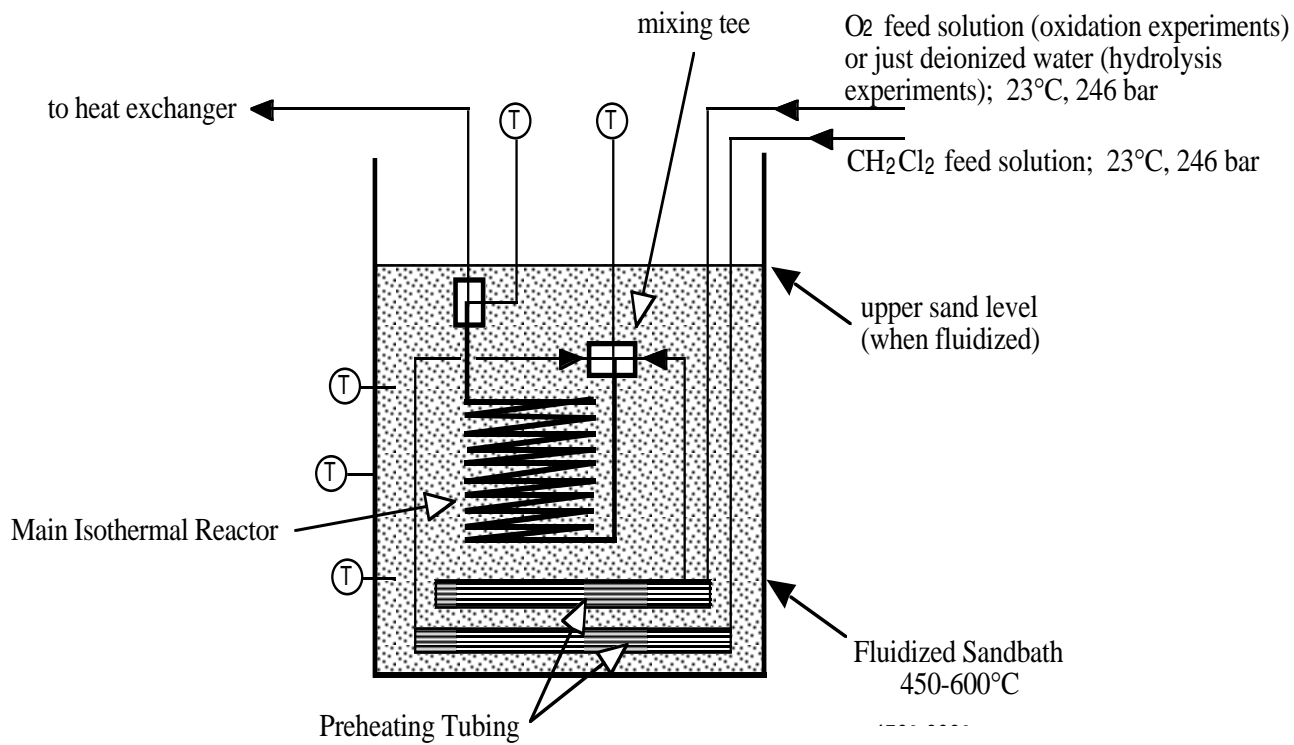

mixing tee

O₂ feed solution (oxidation experiments) or just deionized water (hydrolysis experiments);  23°C, 246 bar

CH₂Cl₂ feed solution;  23°C, 246 bar

to heat exchanger

upper sand level (when fluidized)

Main Isothermal Reactor

Fluidized Sandbath 450-600°C

Preheating Tubing

Ⓣ = thermocouple

**Table A1**

| Tube Orientation | Flow Direction | Flow Type | Correlation Source |
|---|---|---|---|
| Vertical | Upward, Downward | Laminar | Churchill, 1984 [53] |
| Vertical | Upward, Downward | Turbulent | Watts and Chou, 1982 [54] |
| Horizontal | | Laminar | Morcos and Bergles, 1975 [55] |
| Horizontal | | Turbulent | Robakidze *et al.*, 1983 [56] |

# Figure 2

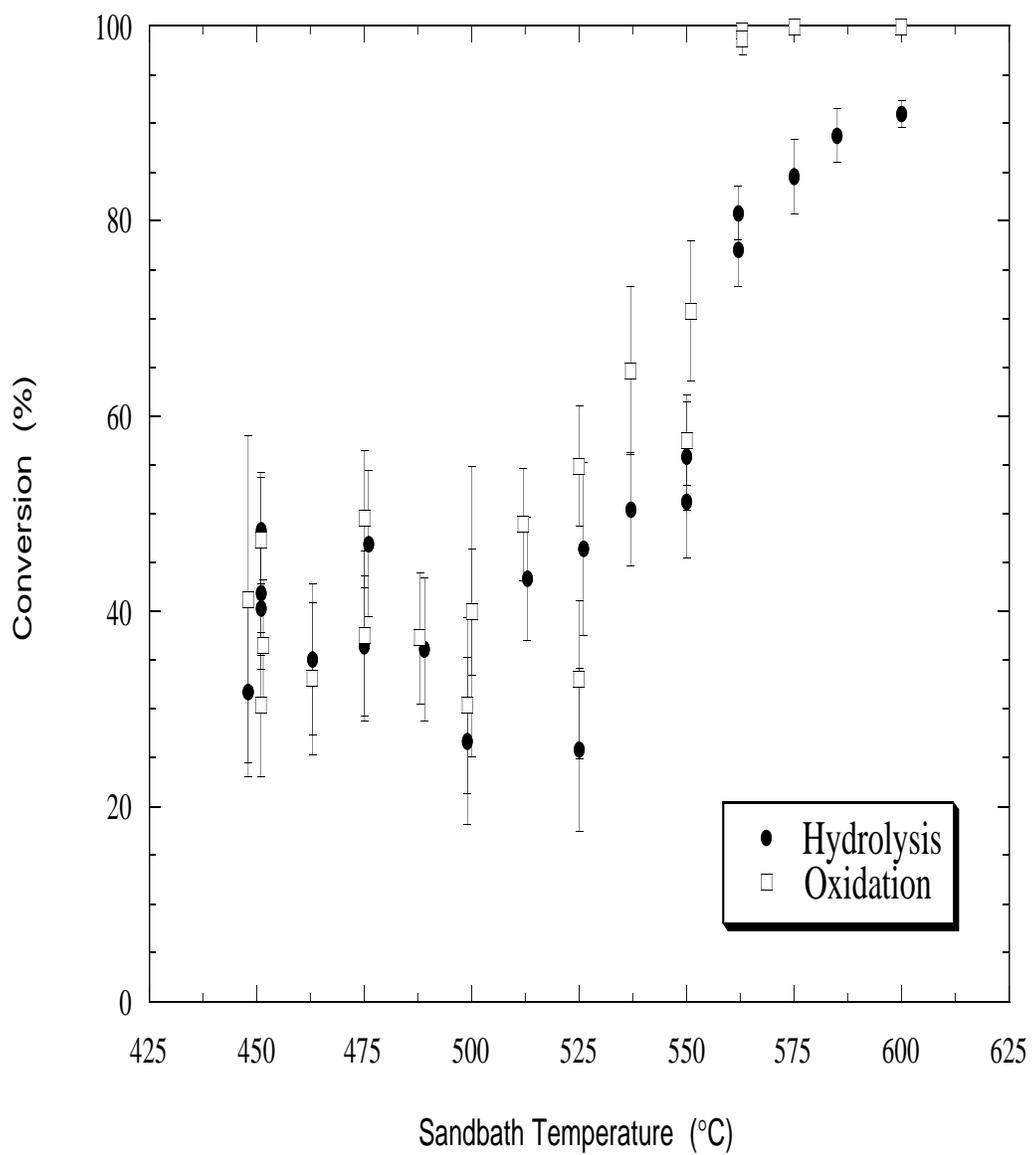

# Figure 3

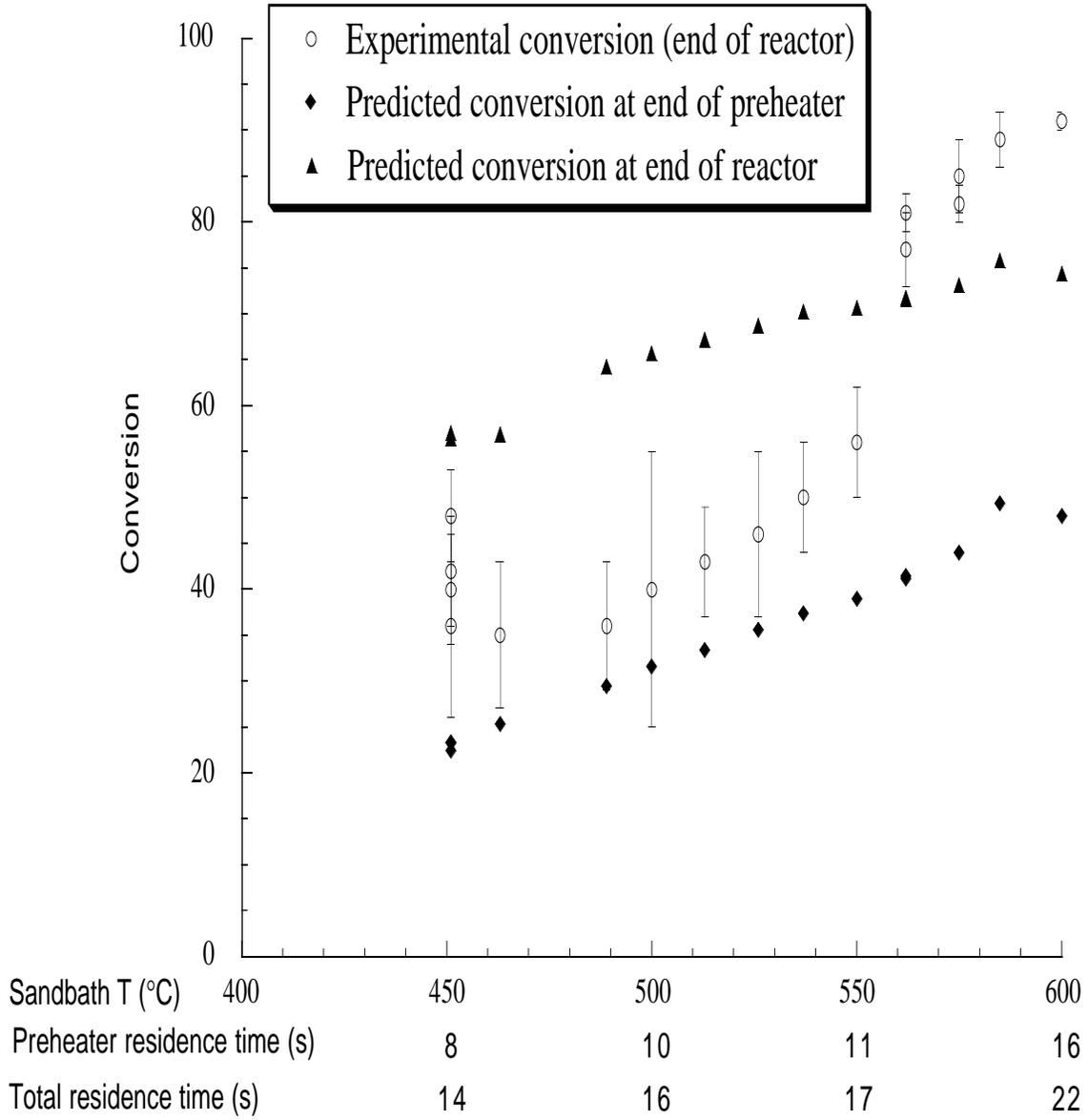



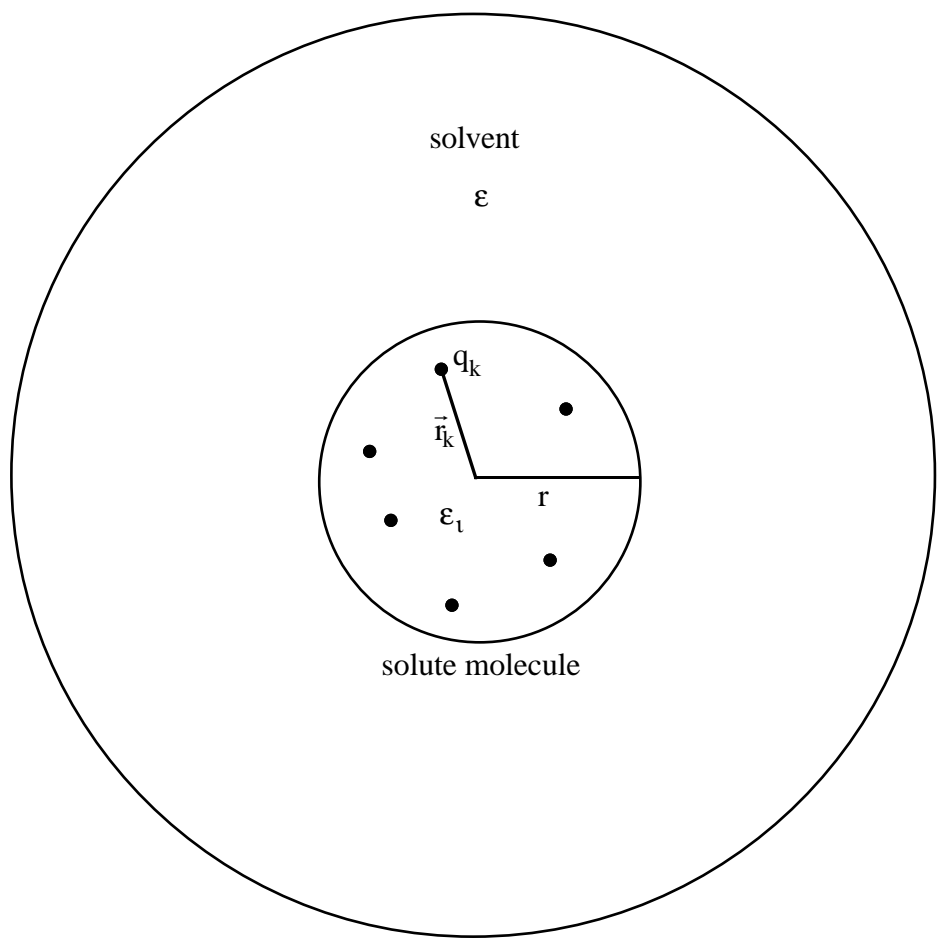

**Table 4**

| TRANSITION STATE PROPERTIES | | | | |
|---|---|---|---|---|

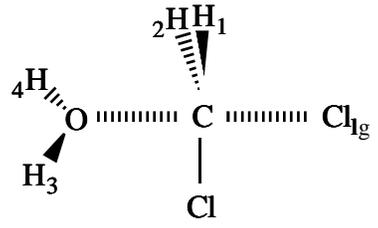

| | | Calculated[a] | Expected[a] | Value in Stable Molecule |
|---|---|---|---|---|
| Bond lengths (Angstroms) | C-O | 1.72 - 1.63 | 1.7 - 1.9 | 1.41 |
| | C-$Cl_{lg}$ | 2.27 - 2.49 | 2.1 - 2.3 | 1.77 |
| | C-$Cl_1$ | 1.78 | | 1.77 |
| | C-$H_{1,2}$ | 1.10 | | 1.09 |
| | O-$H_{3,4}$ | 1.02 | | 0.96 |
| Dipole Moment (Debye) | | 8.21 - 9.95 | | |
| Equivalent Spherical Radius (Angstroms) | | 3.25 - 3.27 | | |

[a] For property values with a range listed, the first value corresponds to that at $\varepsilon = 80$ while the second value corresponds to that at $\varepsilon = 1.3$.

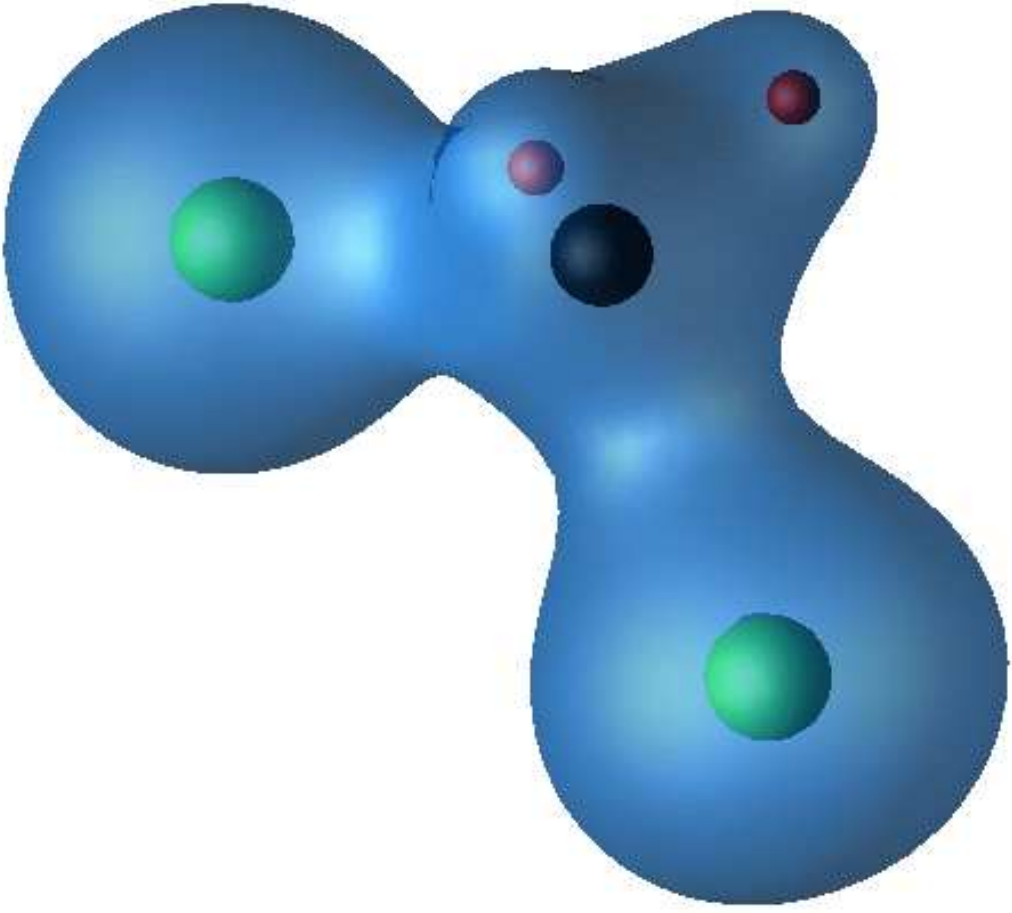

**Table  5**

| Source | Present study | | | | Fells & Moelwyn-Hughes [16] |
|---|---|---|---|---|---|
| T (°C) | 25 | 115 | 374 | 600 | 115 |
| $E_a$ (kJ/mol) | 130 | 131 | 136 | 183 | 107 |

**(a)**

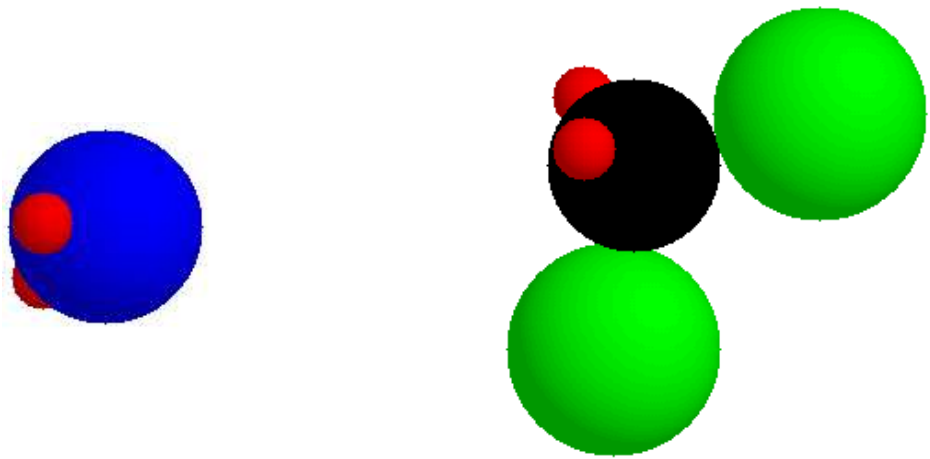

**(b)**

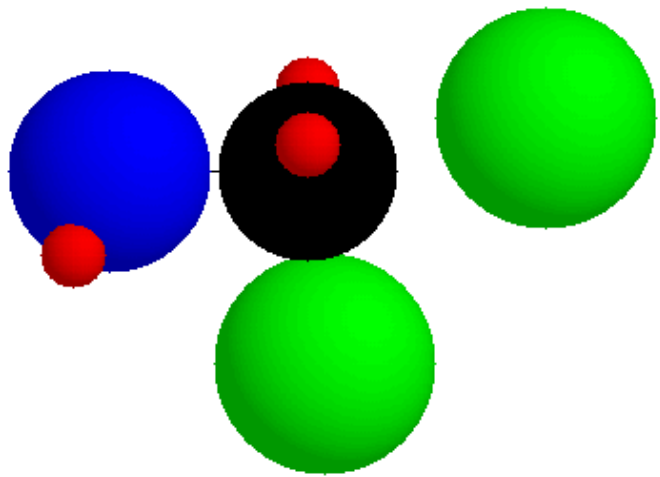

**Table 6**

| T (°C) | 50 | 374 | 550 |
|---|---|---|---|
| $\varepsilon$ | 70.8 | 10.4 | 1.4 |
| $\Phi$ | -0.04 | -1.00 | -7.48 |
| $k_{FMH}\,/\,k'$ (rate decrease factor) | 1.04 | 2.72 | 1765 |

Figure 7

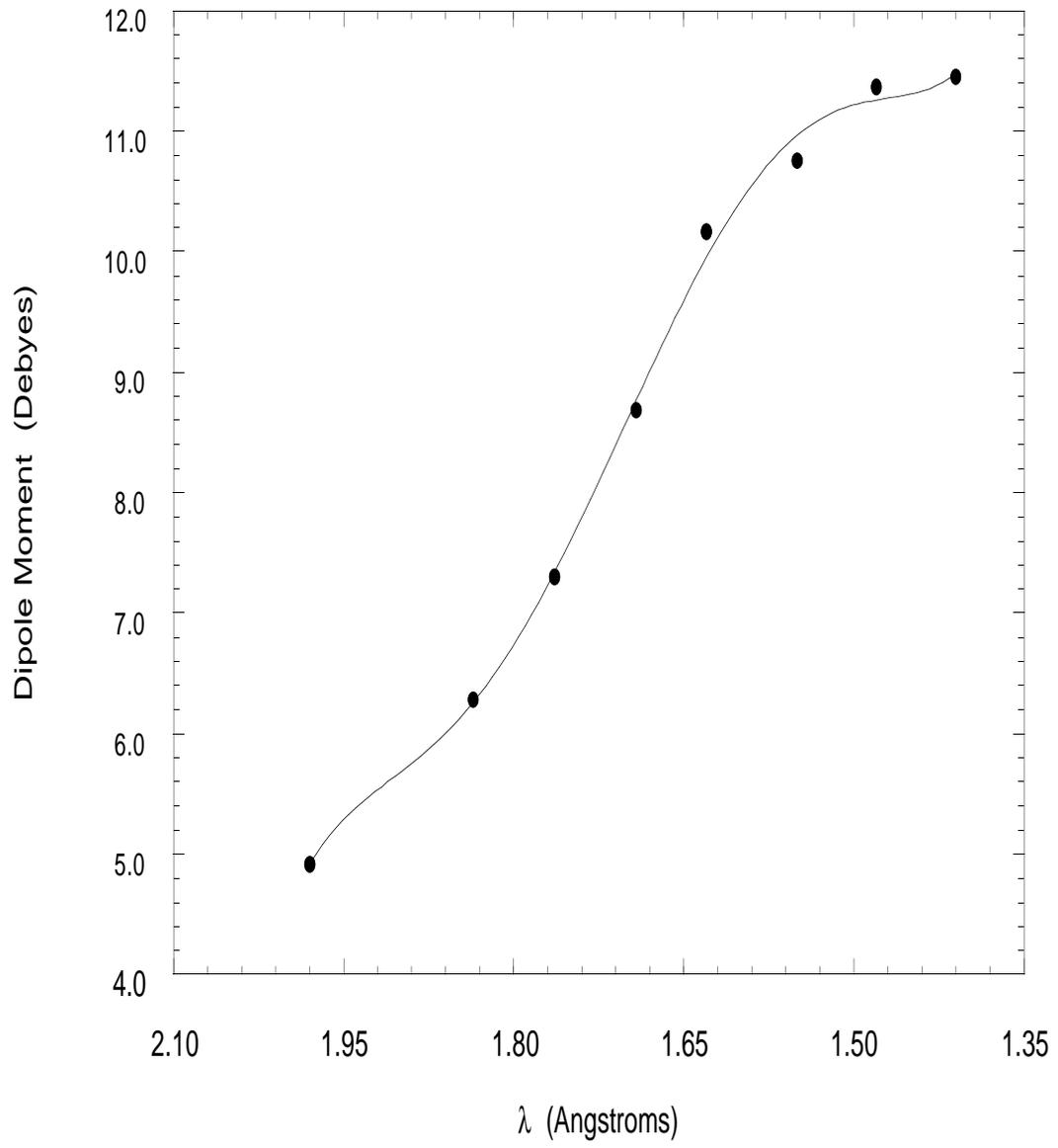

**Table 7**

| $\Phi = \alpha_0 + \alpha_1 T + \alpha_2 T^2 + \alpha_3 T^3 + \alpha_4 T^4 + \alpha_5 T^5 + \alpha_6 T^6 + \alpha_7 T^7 + \alpha_8 T^8 + \alpha_9 T^9$ | | | |
|---|---|---|---|
| Temperature range (°C) | $25 \leq T < 374$ | $374 \leq T \leq 386$ | $386 < T \leq 525$ | $525 < T \leq 600$ |
| $\alpha_0$ | $1.3658 \times 10^{-1}$ | $7.2641851 \times 10^{4}$ | $1.2294119 \times 10^{5}$ | $-2.6826 \times 10^{1}$ |
| $\alpha_1$ | $-1.1319 \times 10^{-2}$ | $-5.8225257 \times 10^{2}$ | $-1.4212424 \times 10^{3}$ | $1.1691 \times 10^{-1}$ |
| $\alpha_2$ | $3.8850 \times 10^{-4}$ | $1.5556709 \times 10^{0}$ | $6.3373575 \times 10^{0}$ | $-2.3416 \times 10^{-4}$ |
| $\alpha_3$ | $-7.9558 \times 10^{-6}$ | $-1.3855168 \times 10^{-3}$ | $-1.2385424 \times 10^{-2}$ | $1.5557 \times 10^{-7}$ |
| $\alpha_4$ | $9.4188 \times 10^{-8}$ | | $4.1293324 \times 10^{-6}$ | |
| $\alpha_5$ | $-6.7546 \times 10^{-10}$ | | $2.3961584 \times 10^{-8}$ | |
| $\alpha_6$ | $2.9717 \times 10^{-12}$ | | $-3.8255570 \times 10^{-11}$ | |
| $\alpha_7$ | $-7.8277 \times 10^{-15}$ | | $1.8223581 \times 10^{-14}$ | |
| $\alpha_8$ | $1.1315 \times 10^{-17}$ | | | |
| $\alpha_9$ | $-6.8988 \times 10^{-21}$ | | | |

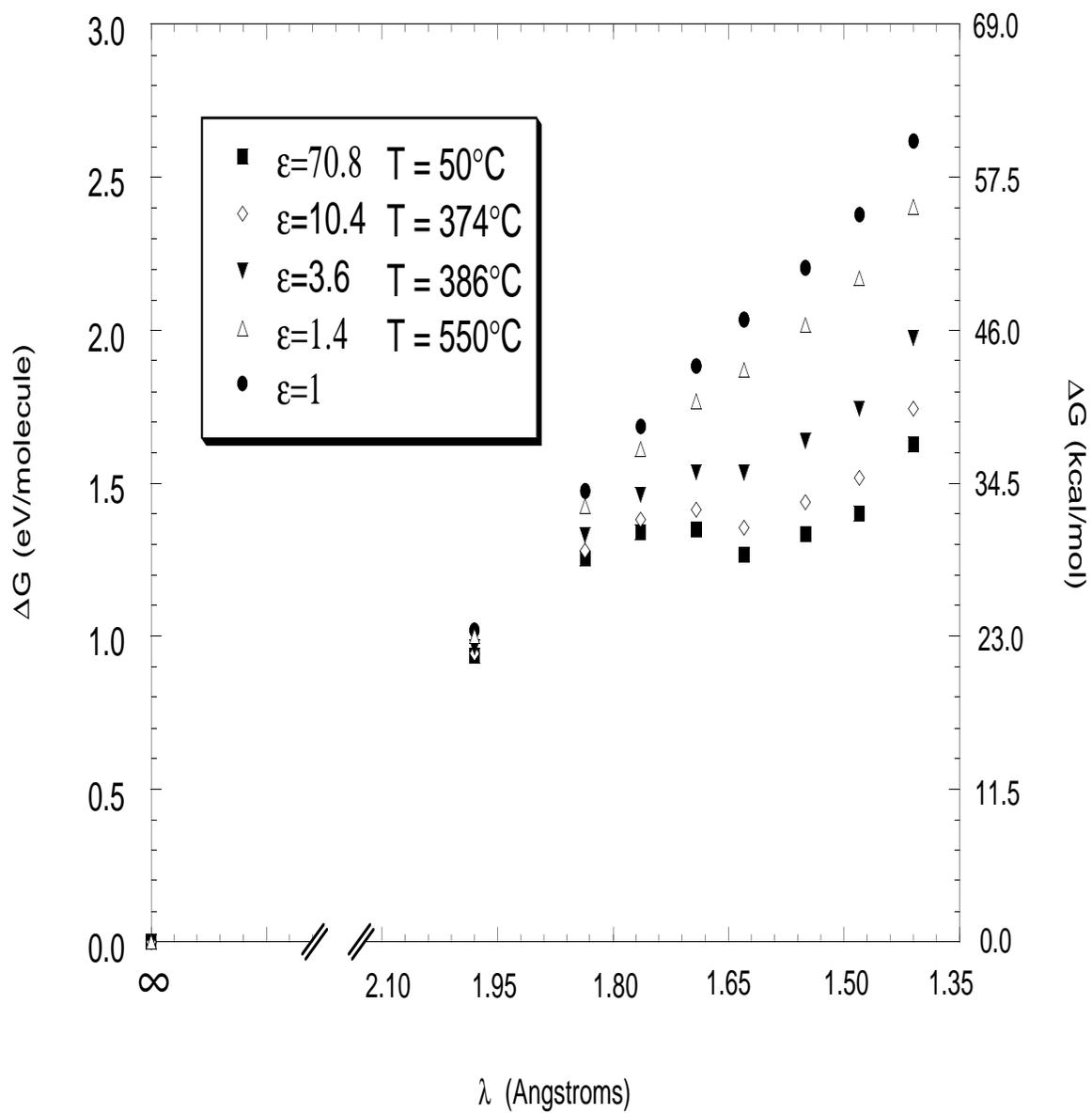

Figure 8

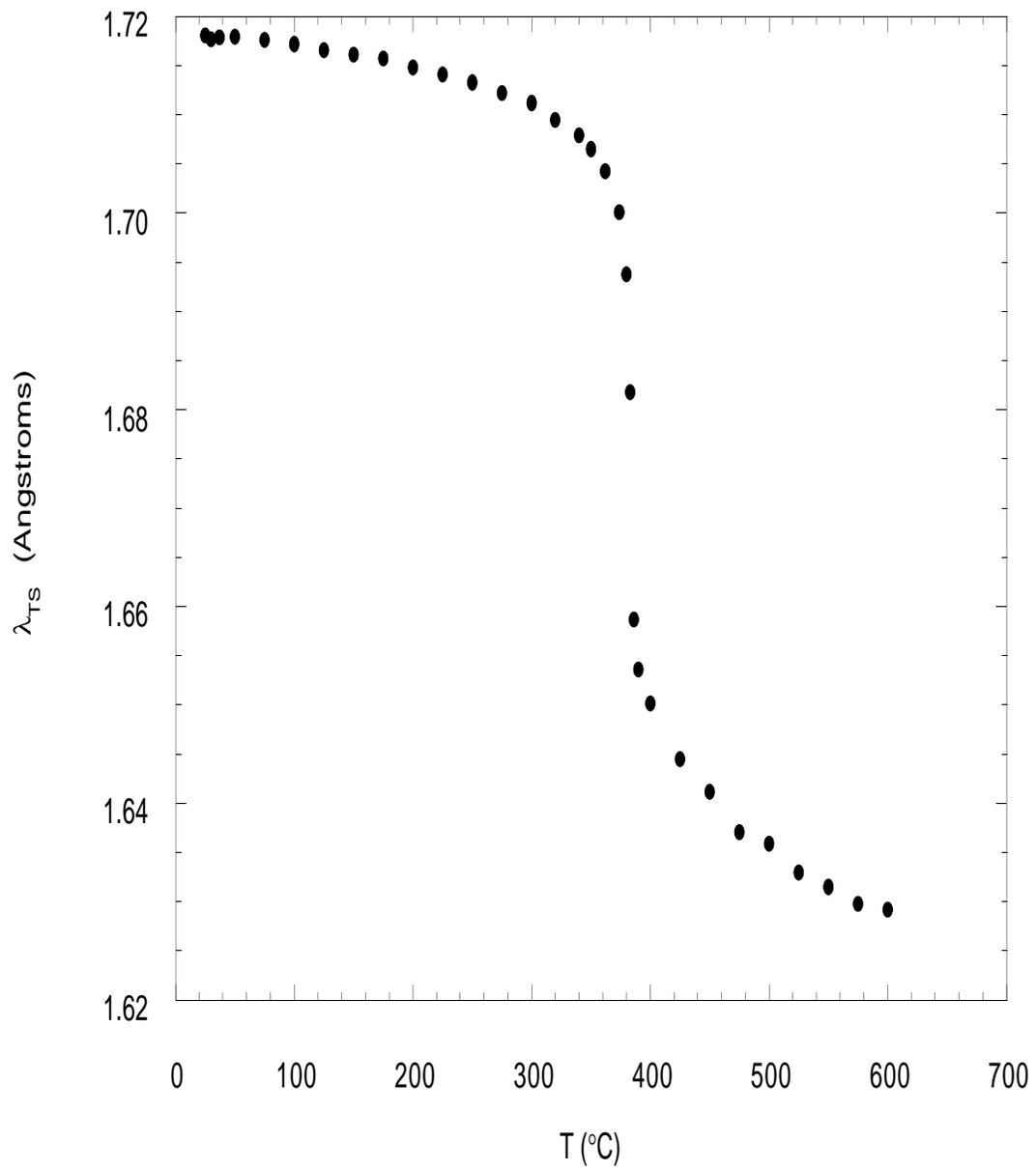

Figure 9

Figure 10

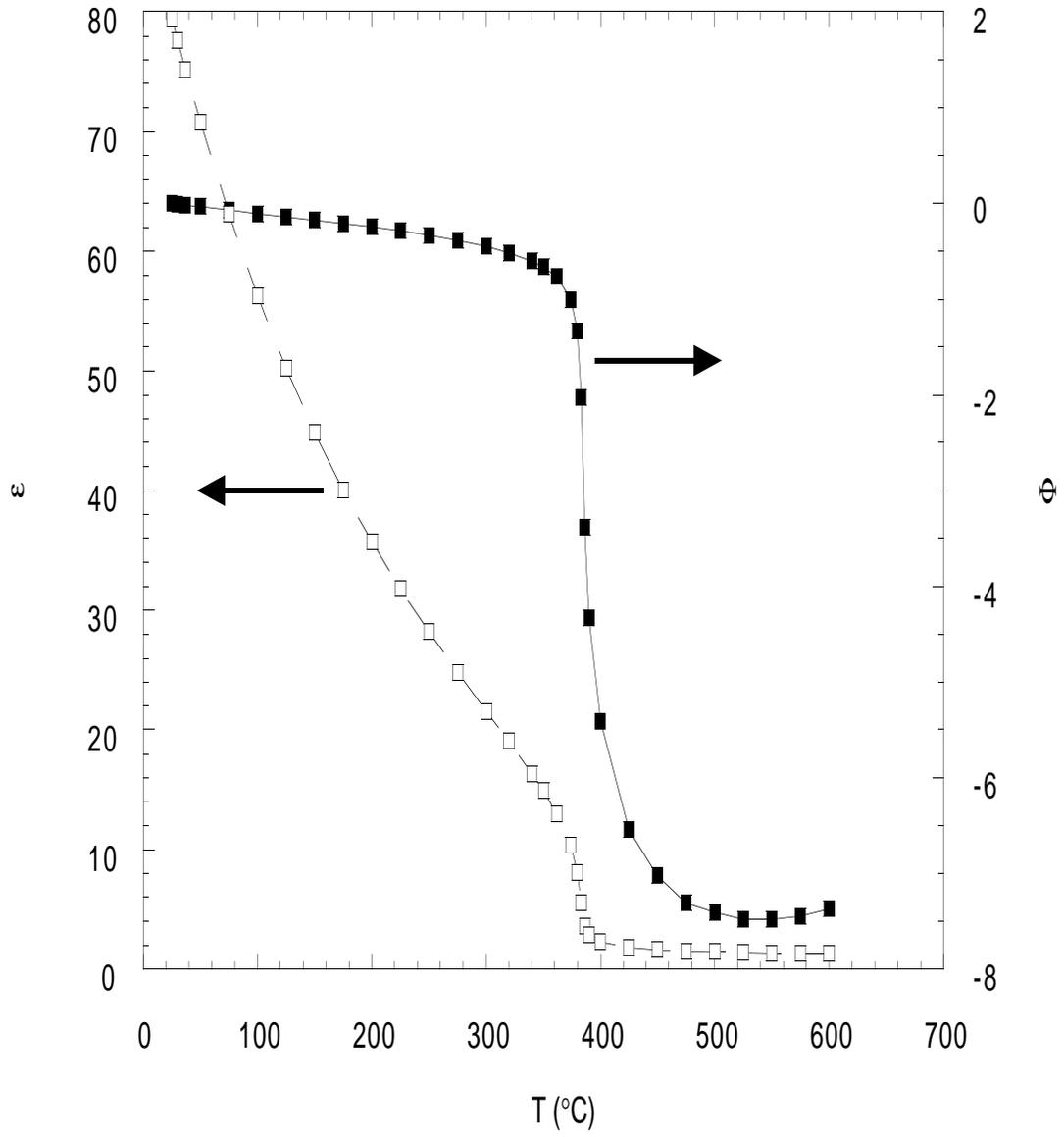

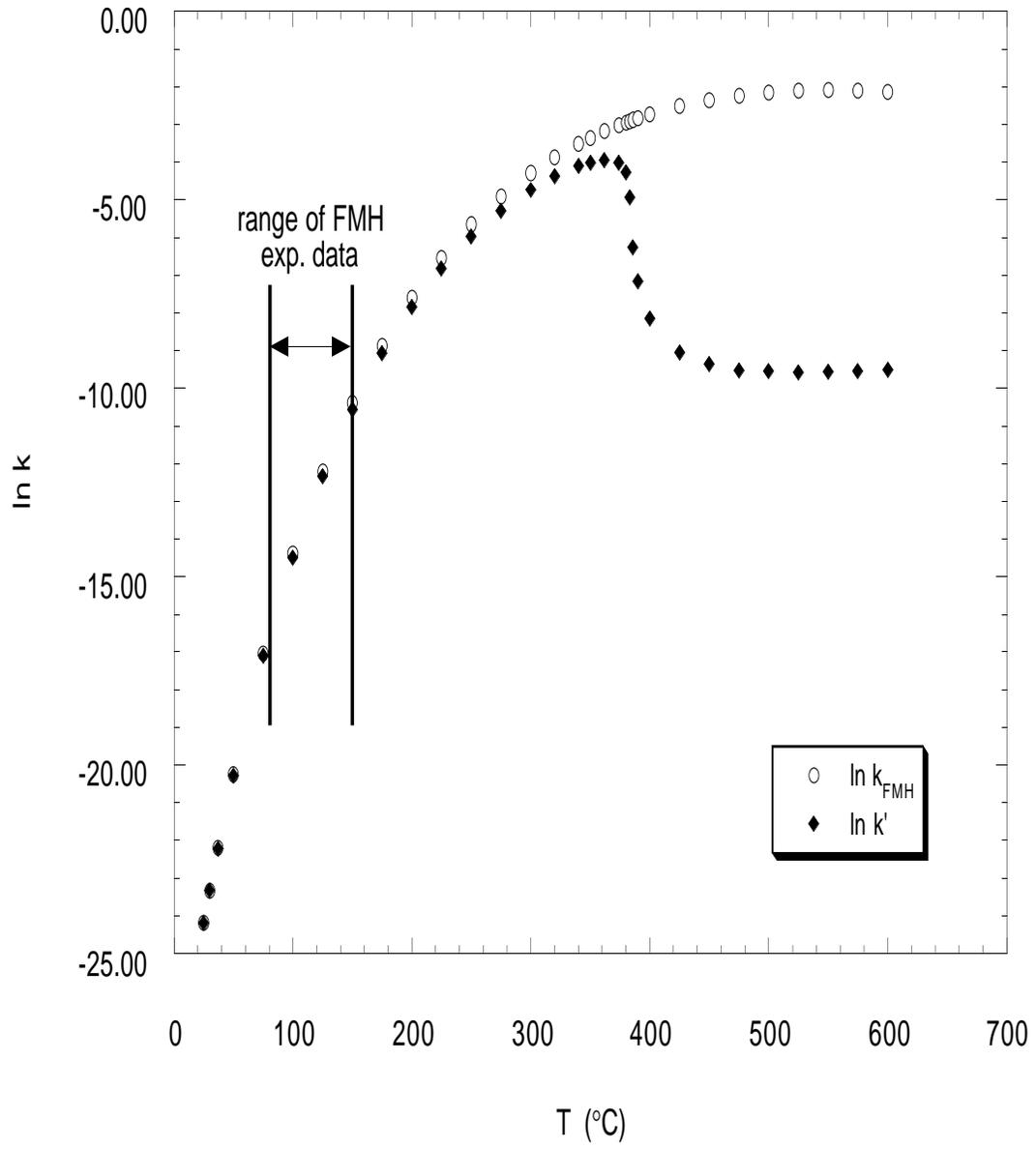

Figure 11

# Figure 12

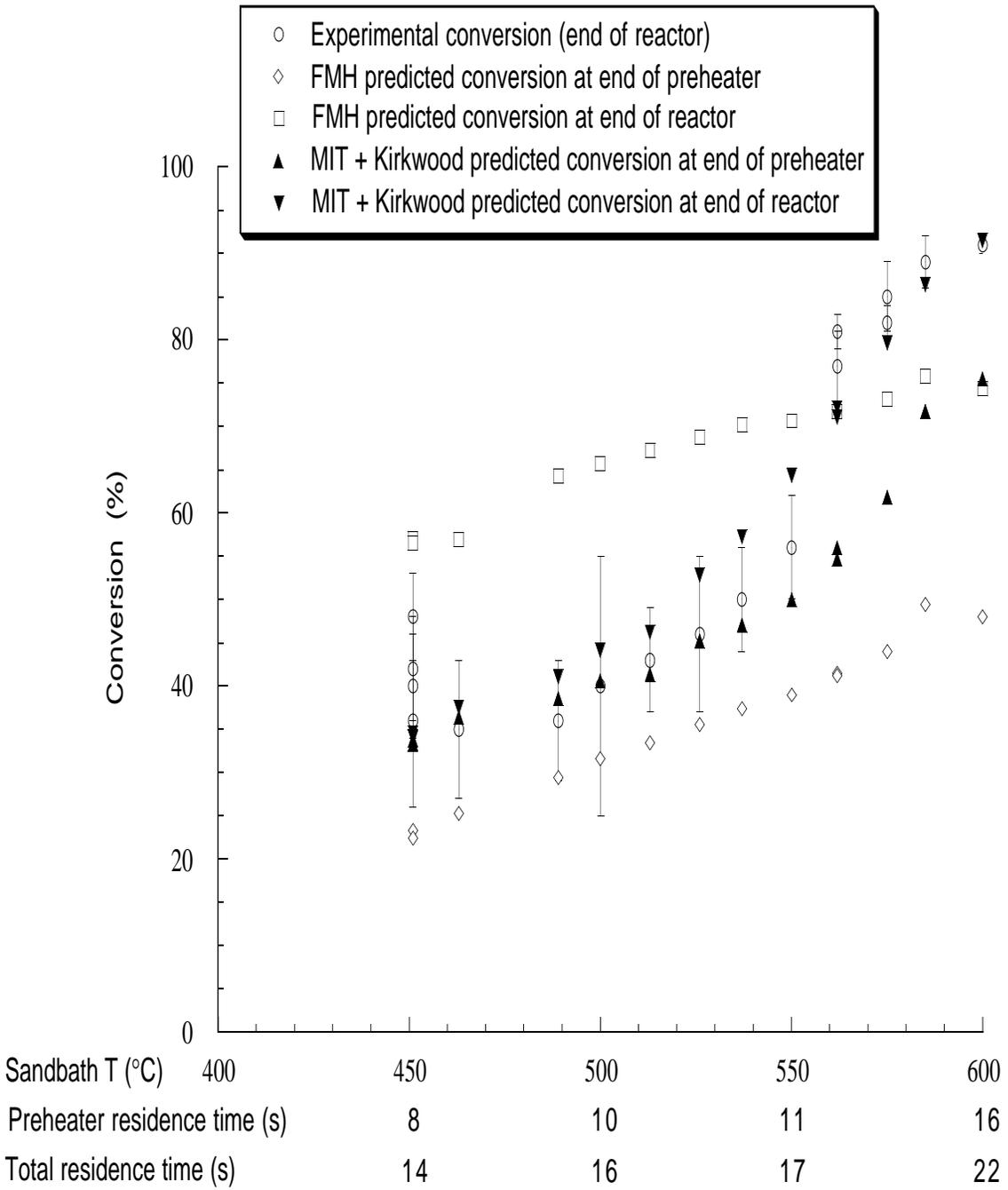

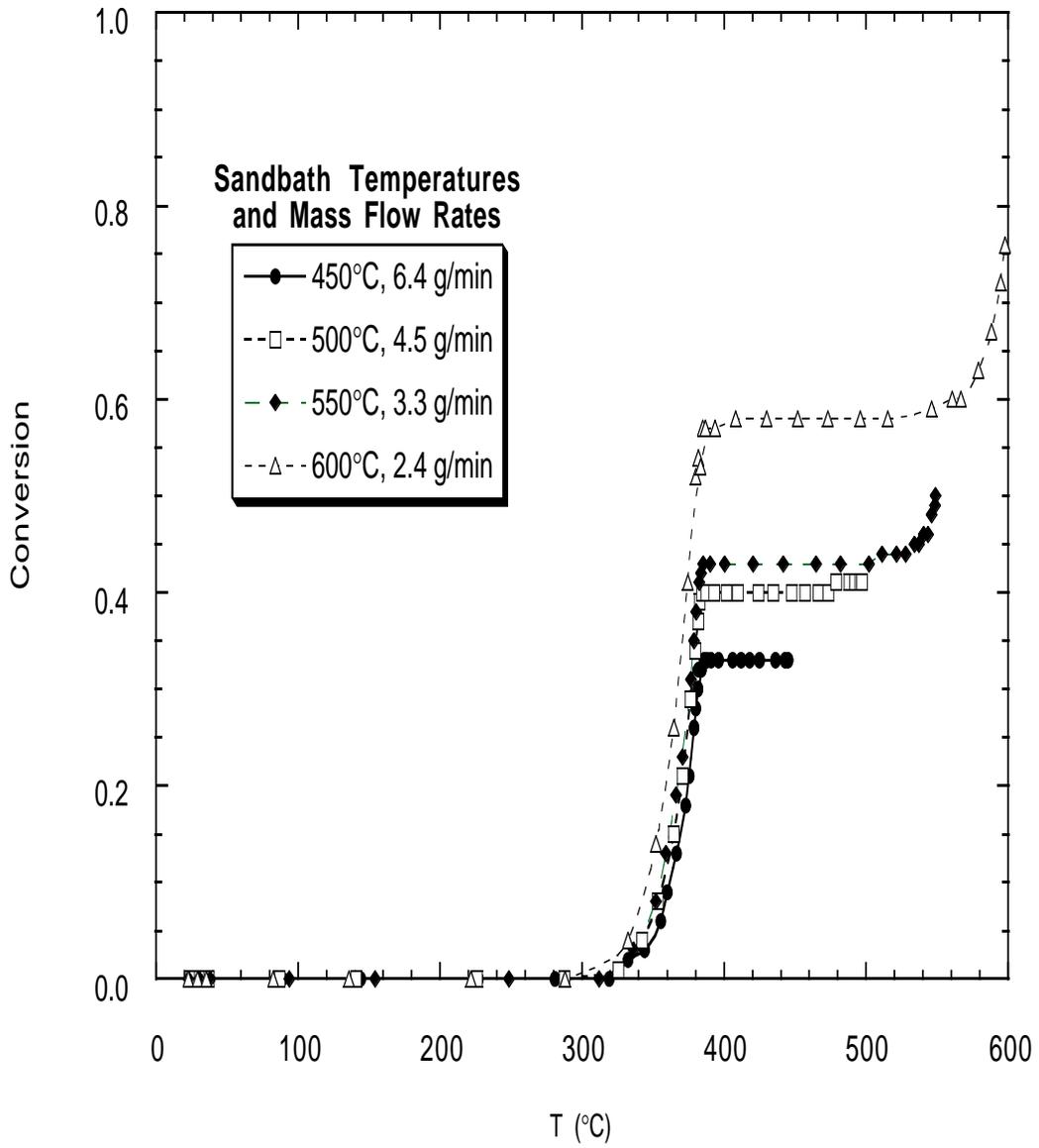

**Predicted conversion vs temperature in preheater**

Sandbath Temperatures
and Mass Flow Rates

— ● — 450°C, 6.4 g/min
-- □ -- 500°C, 4.5 g/min
— ◆ — 550°C, 3.3 g/min
-- △ -- 600°C, 2.4 g/min

Conversion

T (°C)

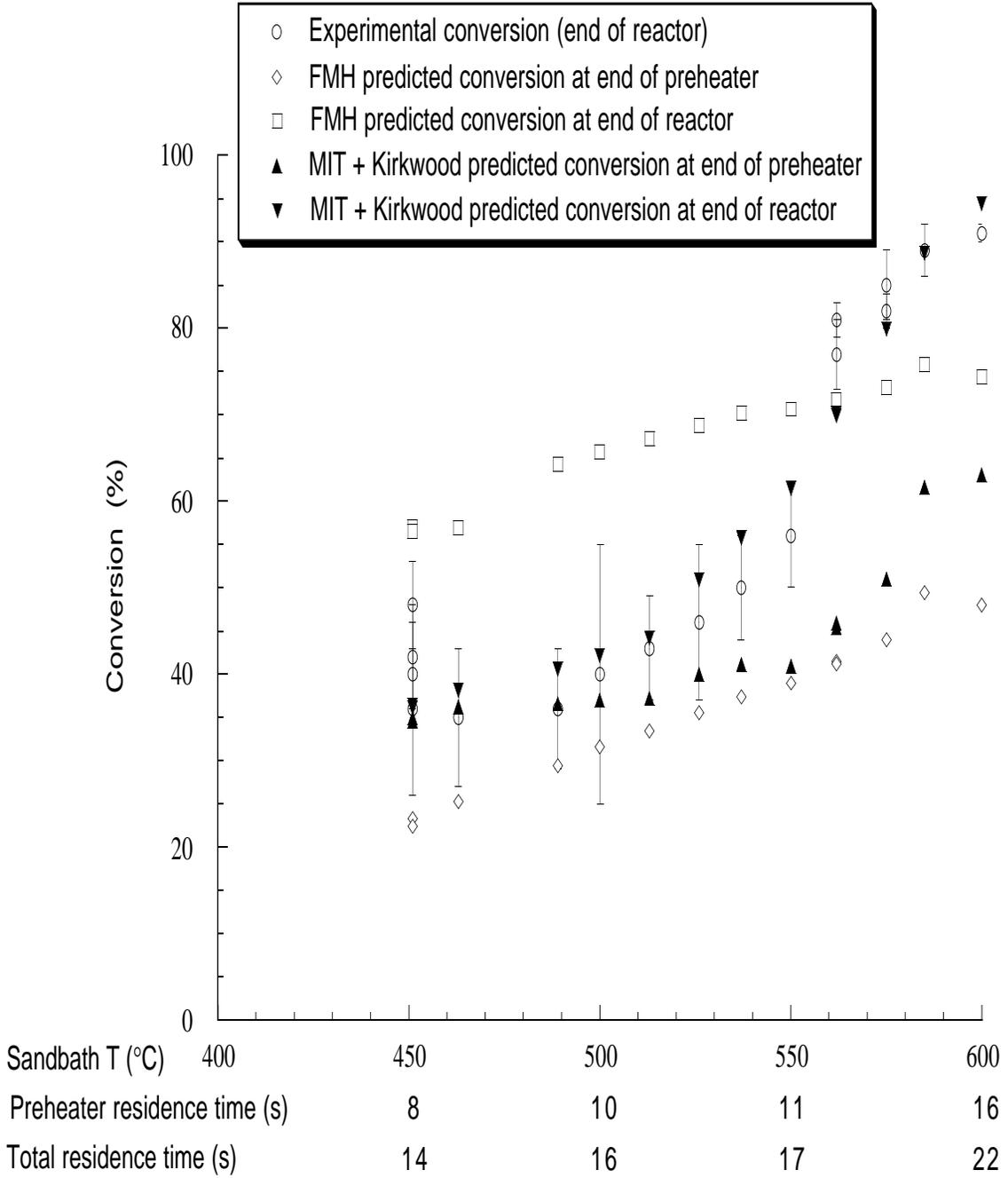

Figure 14

**Table 1**

| | **Preheater Tubing** | **Isothermal Main Reactor** |
|---|---|---|
| Environment | hydrolysis only | hydrolysis + oxidation |
| Temperature | ambient - 600°C | $450°C \leq T \leq 600°C$ |
| Pressure | 246 bar | 246 bar |
| Residence time | 7 - 17 s | 4 - 9 s |
| **Feed Conditions**: | | |
| $[CH_2Cl_2]_o$ | 6 - 38 x $10^{-3}$ mol/L | 0.2 - 0.6 x $10^{-3}$ mol/L |
| $[O_2]_o$ | 14 - 52 x $10^{-3}$ mol/L | 0.56 - 2.08 x $10^{-3}$ mol/L |
| $[O_2]_o$ / $[CH_2Cl_2]_o$ ratio | 0 | 1.0 - 4.9 |
| $CH_2Cl_2$ solution mass flow rate | 2.4 - 6.4 g/min | 2.4 - 6.4 g/min |
| $O_2$ solution mass flow rate | 2.9 - 6.7 g/min | 2.9 - 6.7 g/min |

**Table 2**

| Sandbath Temperature (°C) | CH$_2$Cl$_2$ Conversion (no reactor) | CH$_2$Cl$_2$ Conversion (with reactor; no O$_2$) | Preheater Residence Time (s) [b] | Main Reactor Residence Time (s) [c] |
|---|---|---|---|---|
| 450 | 36±10 % | 40±13 % [a] | 8.1 | 6.0 |
| 575 | 82±2 % | 85±4 % | 13.2 | 6.0 |

[a] Average of four data points

[b] Preheater residence times correspond to fluid temperatures from 25°C up to sandbath temperature

c Main reactor residence times correspond to isothermal fluid temperatures equal to the sandbath temperature

**Table  3**

|  | Calculated | Experimental |
|---|---|---|
| C-H    length | 1.11 Å | 1.09 Å |
| angle | 112.0° | 111.5° |
| C-Cl    length | 1.79 Å | 1.77 Å |
| angle | 112.2° | 112.0° |
| O-H    length | 0.99 Å | 0.96 Å |
| angle | 102.9° | 104.5° |

| Species | Dipole Moment  (Debye) | | Radius  (Angstroms) | | |
|---|---|---|---|---|---|
|  | Calculated | Experimental | vdw (ab initio) | vdw (EOS) | equiv. spherical |
| $H_2O$ | 1.89 | 1.85 | 1.25 | 1.4 | 2.14 |
| $CH_2Cl_2$ | 1.79 | 1.6 | | | 2.98 |